\documentclass[12pt ]{article} 
\usepackage{jheppub_kim}
\usepackage{amsmath}
\usepackage{amssymb}
\usepackage{dcolumn}
\usepackage{bm}
\usepackage{color}
\usepackage{epsfig}
\usepackage{amsfonts}
\usepackage{graphicx}
\usepackage{subfigure}
\usepackage{hyperref}
\begin{document}

\title{Thermodynamics, Shadow, and Quasinormal Modes of AdS  Ay\'{o}n--Beato--Garc\'{i}a  Massive Black Hole}

\author[a,b]{Dharm Veer Singh,\footnote{Visiting Associate: Inter-University Center of Astronomy and Astrophysics (IUCAA), Pune}}
\emailAdd{veerdsingh@gmail.com}

\author[c,d]{Sudhaker Upadhyay,\footnote{Visiting Associate: Inter-University Center of Astronomy and Astrophysics (IUCAA), Pune}}
\emailAdd{sudhakerupadhyay@gmail.com}

\author[a]{Amit Kumar,}
\emailAdd{ammiphy007@gmail.com}

\author[e]{Yerlan Myrzakulov,}
\emailAdd{ymyrzakulov@gmail.com}

\author[e]{Kairat Myrzakulov}
\emailAdd{krmyrzakulov@gmail.com}
 \author[f]{Himanshu Kumar Sudhanshu}
\emailAdd{himanshu4u84@gmail.com}

\affiliation[a]{Department of Physics, Institute of Applied Science and 
Humanities, GLA University, Mathura - 281406, Uttar Pradesh, India}
\affiliation[b]{School of Physics, Damghan University, P.O. Box 
3671641167, Damghan, Iran}
\affiliation[c]{Department of Physics, K.L.S. College, Magadh University, 
Nawada 805110, India}
\affiliation[d]{School of Physics, Damghan University, P.O. Box 
3671641167, Damghan, Iran}
\affiliation[e]{Department of General \& Theoretical Physics, L. N. 
Gumilyov Eurasian National University, Astana, 010008, Kazakhstan}
 \affiliation[f]{Department of Physics, J.J. College, Gaya, Magadh University, Bodh Gaya, Bihar 823003, India}  
\abstract{
We investigate the thermodynamics, photon sphere, and dynamical stability of an AdS Ay\'{o}n--Beato--Garc\'{i}a (ABG) massive black hole with graviton mass and magnetic charge. 
The Gibbs free energy exhibits distinct limiting behaviors: it reduces to that of an AdS massive black hole when magnetic charge vanishes, to that of an AdS ABG black hole when graviton mass is zero, and smoothly interpolates to the AdS massive Reissner–Nordstr\"om case in the asymptotic regime. Furthermore, the photon sphere and shadow analysis indicate that increasing the graviton mass expands their radii, while increasing the magnetic charge causes contraction, in agreement with earlier studies of black hole spacetimes. Quasinormal mode (QNM) calculations further confirm dynamical stability, as the imaginary part remains negative, ensuring decay of perturbations. Additionally, the real part of the frequency decreases with graviton mass, while the imaginary part initially grows before saturating at higher values. Together, these results provide meaningful insights into the interplay between graviton mass, magnetic charge, and stability, thereby enriching the understanding of black holes in modified gravity theories.
}

\maketitle
%%%%%%%%%%%%%%%%%%%%%%%%%%%%%%%%%%%%%%%%%%%%%%%%%%%%%%%%%%%%%%%%%%%%%
\section{Introduction}\label{sec:level1}
Black holes represent one of the most intriguing theoretical predictions from General Relativity  (GR). Numerous observational studies have been conducted to validate their presence. The first definitive confirmation of their existence was achieved when the Laser Interferometer Gravitational Wave Observatory (LIGO) detected gravitational waves generated by a black hole merger \cite{lig}.
 However, GR is incomplete, and theory remains a significant challenge in contemporary scientific research. This difficulty arises from the fact that GR, while highly successful in describing gravitational interactions, is not a theoretically complete framework.  
Numerous researchers have undertaken diverse efforts to extend the GR by introducing modifications to Einstein's theory of gravity. These attempts address its limitations and provide a more comprehensive framework for describing gravitational interactions.
An alternative approach to modifying the GR involves incorporating massive gravitons, a concept that has garnered significant interest within the scientific community. A consistent theory of massive gravity that avoids ghost instabilities has been formulated, as outlined in Ref. \cite{17}. However, such modifications may introduce ghost-like instabilities when extended to curved spacetime, as discussed in Ref. \cite{18}. The implications of massive gravity on black hole solutions have also been explored, particularly concerning its influence on spacetime geometry \cite{19, 20}.  

Born and Infeld introduced nonlinear electrodynamics (NLED) to resolve point charge singularities and energy divergence \cite{4}. NLED gained prominence as it emerged in certain limits of string theory \cite{5}. Because classical electrodynamics fails at high energies due to interactions with other fields, NLED is a viable alternative. As a gravitational source, NLED generates diverse black hole solutions \cite{6,7,8,9} and plays a crucial role in cosmology \cite{10,11,12} and string theory \cite{5,13}. ABG proposed the first black hole solution with an NLED field satisfying the weak energy condition \cite{14}.

    The study of black hole thermodynamics was initially pioneered by Bekenstein \cite{49, 50} and Hawking \cite{51}, who formulated a fundamental relationship between entropy and the area of the event horizon of a black hole. Subsequent research has demonstrated that black hole entropy is directly proportional to the surface area of its horizon \cite{52, 53}. Over time, the thermodynamic properties of black holes have continued to attract considerable academic interest, leading to extensive investigations from various perspectives \cite{54,55,56,57,58,59,60,61,62}. The size and morphology of a black hole shadow serve as key indicators of its fundamental properties, such as mass, spin, and spacetime geometry \cite{029,030,031,032,033}. Analyzing the structure of the shadow enhances our understanding of gravity under extreme conditions. Recent studies have examined the shadows of the Reissner-Nordstr\"om BH \cite{034}, the charged massive BTZ black hole \cite{035}, and black holes coupled with  NLED \cite{036}. Additionally, BH shadows are closely linked to quasinormal modes (QNMs) \cite{037, 038}, which characterize perturbation dynamics. QNMs of regular black holes have been investigated in \cite{041}.

 In this work, we explore the exact black hole solution in the context of NLED within massive gravity. The gravitational action incorporating the Ricci scalar, massive graviton terms, and the NLED Lagrangian is introduced, leading to the field equations derived from metric variation. We detail the Einstein tensor and massive gravity contributions, with explicit expressions for the auxiliary polynomials and reference metric. Using a spherically symmetric Ansatz, we derive the black hole metric function for an ABG-like source. The solution's asymptotic behaviour aligns with that of the massive Reissner-Nordström black hole, and its event horizon structure is analysed numerically.
Furthermore, we discuss the dependence of the horizon on the black hole’s mass, magnetic charge, and graviton mass. We also examine the thermodynamic properties, including mass, temperature, and entropy, highlighting deviations from the standard area law. We find that modifying the first law of thermodynamics is necessary to ensure consistency with expectations of entropy. We assess stability through heat capacity and Gibbs free energy, which indicate phase transitions and thermodynamic equilibrium conditions.

 The structure of the paper is as follows. In Sec. \ref{sec2}, we derive the field equations through metric variation, examine the contributions of massive gravity terms, and establish key assumptions underlying the black hole solution. We analyze the physical significance of parameters such as mass, charge, graviton mass, and AdS length, explore the asymptotic behaviour of the solution, compare it with known black holes, and investigate the event horizon structure using numerical and graphical methods. In Sec. \ref{sec3}, we determine the black hole mass from the horizon condition and compute its temperature and entropy. Additionally, we discuss deviations from the standard area law, examine correction factors, and assess the stability of the black hole. In Sec. \ref{sec4}, we analyze the photon radius and black hole shadow by deriving the equations of motion and studying null circular geodesics. The effects of the graviton mass and magnetic charge on the photon sphere and shadow radius are examined through numerical and graphical results. In Sec. \ref{sec5}, we investigate  QNMs to evaluate black hole stability, derive the wave equation, apply the WKB approximation to obtain  QNFs, and analyze their dependence on black hole parameters. Finally, we present our conclusions, summarising the key findings and highlighting the impact of modified gravity on black hole thermodynamics.
\section{Exact solution of ABG  black hole with massive gravity}\label{sec2}
Let us explore gravity, incorporating an NLED source within the framework of massive gravity. The corresponding action is expressed as follows:
\begin{equation}
S=\int d^4x \sqrt{-g}\left[R-2\Lambda+m^2\sum_{i=1}^{2}c_{i}\,\mathcal{U}_{i}(g,f)+\mathcal{L} (F)\right],
\label{action}
\end{equation}
where $R$ denotes the Ricci scalar curvature, $m$ the graviton mass, $c_i$ the coefficients of the massive gravity terms, and $f$ the symmetric tensor. 
The function $\mathcal{L}(F)$, representing the Lagrangian density in NLED, is defined in terms of the Faraday tensor $F$, given by $F = F_{\mu \nu } F^{\mu \nu}$, where $F_{\mu \nu }$ is the electromagnetic field tensor. The considered system is obtained by
using the  Legendre transformation ($H=2F\partial L/\partial F-L$), which depends on the antisymmetric field ($P = P_{\mu \nu } P^{\mu \nu}$). Thus, the Lagrangian can be expressed as
\begin{equation}
    L=2PH_P-H, \qquad \text{and}\qquad H_P=\frac{\partial H}{\partial P},
\end{equation}
where $F_{\mu \nu }=H_P P_{\mu \nu }$ is the electromagnetic field and $H(P)$ is the structural function corresponding to $F_{\mu \nu }$.
-
To derive the equations of motion, the action (\ref{action}) is varied with respect to the metric tensor ($g_{\mu \nu}$) and the electromagnetic potential ($A_{\mu}$), leading to the following expressions:
\begin{eqnarray}
G_{\mu \nu }+\Lambda g_{\mu\nu}+m^{2}\varrho _{\mu \nu }&=&T_{\mu\nu}  \equiv2\left[H(P) P_{\mu\rho}P^{\rho}_{\nu}- 2P H_P+H(P)\right],\\
\nabla_{\mu}( P^{\mu\nu})=0.
\label{Field equation}
\end{eqnarray}
Here, $G_{\mu\nu}$ signifies the Einstein tensor and 
$\varrho_{\mu\nu}$   encapsulates the contribution from the massive gravity framework with the following explicit expressions:
\begin{eqnarray}
G_{\mu\nu}&=&R_{\mu\nu}-\frac{1}{2}g_{\mu\nu}R,\\
\varrho _{\mu \nu }& =&-\frac{c_{1}}{2}\left( \mathcal{U}_{1}g_{\mu \nu }-
\mathcal{K}_{\mu \nu }\right) -\frac{c_{2}}{2}\left( \mathcal{U}_{2}g_{\mu
\nu }-2\mathcal{U}_{1}\mathcal{K}_{\mu \nu }+2\mathcal{K}_{\mu \nu
}^{2}\right). 
\label{eom1}
\end{eqnarray} 
where $\mathcal{U}_{1}$ and  $\mathcal{U}_{2}$ are polynomials of eigenvalues of the matrix $\mathcal{K}_{\nu }^{\mu }=\sqrt{g^{\mu \alpha }f_{\alpha \nu }}$ expressed as   \cite{Cai:2014znn}
\begin{eqnarray}
\mathcal{U}_{1} &=&\left[ {K}\right] ,\\
\mathcal{U}_{2} &=&\left[ {K}\right] ^{2}-\left[ {K}^{2}
\right],
\end{eqnarray}
with $\left[{K}\right]= {K}_{\mu }^{\mu }$.   The reference metric is assumed to take the following form, as proposed in \cite{Cai:2014znn}:
\begin{equation}
f_{\mu \nu }=\text{diag}(0,0,c^{2}h_{ij}).
  \label{f11}
\end{equation}
For this reference metric, the polynomials  $\mathcal{U}_{1}$ and  $\mathcal{U}_{2}$  read
\begin{eqnarray}
&&\mathcal {U}_1 =\frac{2c}{r},\qquad \mathcal {U}_2=\frac{2c^2}{r^2}.
\end{eqnarray}
To determine the black hole solution, we employ a static, spherically symmetric spacetime, represented by the following line element:
\begin{equation}
ds^2=-f(r)dt^2 +\frac{1}{f(r)}dr^2+r^2d\Omega^2, \qquad\text{with}\qquad f(r)=1-\frac{2{m(r)}}{r}.
\label{m2}
\end{equation}
 The ABG black hole originates from the following structural function ($H(P)$) of the NLED source, 
 given by \cite{abg00, Singh:2021nvm}:
\begin{equation}
{H(P)}= \frac{F(1-(\beta-1)\sqrt{-2g^2P})}{(1+\sqrt{-2g^2P})^{1+\beta/2}}-\frac{\alpha}
{-2g^2s}\left(\frac{(2g^2P)^{5/4}}{(1+\sqrt{-2g^2P})^{1+\alpha/2}}\right),
\label{nonl1}
\end{equation}
where $\alpha$ and $\beta$ are the dimensionless parameters, when $\alpha=3$ and $\beta=4$, it generalises to ABG source.

In the regime of the weak field approximation, the Lagrangian density (\ref{nonl1}) reduces to that of Maxwell electrodynamics, where ${H(P)}=P$ for the weak fields ($P<<1$), and simultaneously adheres to the weak energy condition, ensuring $H<0$ and $H_P>0$ \cite{Singh:2022xgi}. The model is characterised by three free parameters: $ s = g/2M$, $M$, and $g$, which are associated with the magnetic charge and mass. The sole non-zero component of the electromagnetic field tensor $F_{\mu\nu}$ is given by $F_{\theta\phi}=g(r)\sin\theta$, with the corresponding electromagnetic potential expressed as $A_\phi=-g(r)\cos\theta$ \cite{Singh:2022xgi,Kumar:2024cnh}. { The non-zero components of the energy-momentum tensor are given
by \cite{Kumar:2024cnh}
\begin{eqnarray}
  &&  T^t_t=T^r_r=\frac{g^2(3g^2-r^2)}{2(r^2+g^2)^3}- \frac{3M g^2}{(r^2+g^2)^{5/2}},\nonumber\\
 &&   T^\theta_\theta=T^\phi_\phi= \frac{g^2(3g^4-8g^2r^2+r^4)}{(r^2+g^2)^4}+\frac{3g^2M(2g^4-g^2r^2-3r^4)}{(r^2+g^2)^{9/2}}.
\end{eqnarray}

By plugging the expression for $f(r)$  into the $(r,r)$ component of Eq. (\ref{eom1}), the resulting $(r,r)$ components of the field equations, as represented in Eq. (\ref{Field equation}), are given by
\begin{eqnarray}
&&-m'(r)-\frac{m^2}{2}\left({cc_1 r}+c^2c_2\right)-\frac{3 r^2}{2l^2}=\frac{g^2r^2(3g^2-r^2)}{2(r^2+g^2)^3}- \frac{3M g^2r^2}{(r^2+g^2)^{5/2}},
\label{em}
\end{eqnarray}
where the prime denotes the first derivative of $r$. Integrating Eq. (\ref{em}) in the limit $r$ to $\infty$, we get
\begin{equation}
  -  m(r)- \frac{m^2}{2}\left(\frac{cc_1 r^2}{2}+c^2c_2r\right)-\frac{r^3}{2l^2}+C_1=\int_r^\infty\left(\frac{g^2r^2(3g^2-r^2)}{2(r^2+g^2)^3}- \frac{3M g^2 r^2}{(r^2+g^2)^{5/2}}\right)dr,
\end{equation}
where
\begin{equation}
    C_1=\lim_{r\to\infty}\left(  m(r)+ \frac{m^2}{2}\left(\frac{cc_1 r^2}{2}+c^2c_2r\right)+\frac{r^3}{2l^2}\right) =M,
    \label{em1}
\end{equation}
and
\begin{equation}
\int_r^\infty\left(\frac{g^2r^2(3g^2-r^2)}{2(r^2+g^2)^3}- \frac{3M g^2 r^2}{(r^2+g^2)^{5/2}}\right)dr=M- \frac{M r^3}{(r^2+g^2)^{3/2}}+\frac{g^2r^3}{2(r^2+g^2)^2}.
\label{em2}
\end{equation}
Here, $C_1$ represents an integration constant, which can be interpreted as the mass of the black hole. Substituting Eqs. (\ref{em1}) and (\ref{em2}) in Eq. (\ref{em}), $m(r)$ reads
\begin{equation}
    m(r)=\frac{M r^3}{(r^2+g^2)^{3/2}}-\frac{g^2r^3}{2(r^2+g^2)^2}-\frac{r^3}{2l^2}-\frac{m^2}{2}\left(\frac{cc_1 r^2}{2}+c^2c_2r\right).
\end{equation}
The corresponding black hole metric is expressed as:
\begin{eqnarray}
f(r)=1-\frac{2M r^2}{(r^2+g^2)^{3/2}}+\frac{g^2r^2}{(r^2+g^2)^2}+m^2\left (c^2c_2+\frac{cc_1 r}{2}\right)+\frac{r^2}{l^2}.
\label{bhs}
\end{eqnarray}
}
This black hole solution (\ref{bhs}) is characterized by different parameters, such as mass ($M$), magnetic charge ($g$), graviton mass ($m$), and the $AdS$ length related to the cosmological constant $(l=\sqrt{-3/\Lambda})$. {Here  $c_1$ and $c_2$ are the massive gravity parameter.  The solution (\ref{bhs}) interpolates with the massive black hole \cite{bb15} in the absence of magnetic charge ($g=0$), $AdS$ ABG black hole in the absence of graviton mass \cite{Singh:2021nvm} ($m=0$) and Schwarzschild black hole in the absence of graviton mass ($m=0$) and magnetic charge ($g=0$). It has been observed that the derived black hole solution (\ref{bhs}) exhibits asymptotic behaviour consistent with that of a Reissner-Nordstr\"om massive black hole
\begin{equation}
f(r)=1-\frac{2M}{r}+\frac{g^2}{r^2}+m^2\left (c^2c_2+\frac{cc_1 r}{2}\right)+\frac{r^2}{l^2}+O\left(\frac{1}{r^3}\right).
\end{equation}
The event horizon of a black hole is determined by the condition  $f(r)=0$, i.e., 
  \begin{equation}
1-\frac{2M r^2}{(r^2+g^2)^{3/2}}+\frac{g^2r^2}{(r^2+g^2)^2}+m^2\left (c^2c_2+\frac{cc_1 r}{2}\right)=0.
\label{eq.hor}
  \end{equation}
The black hole horizon depends upon parameters $M,g,m$, and $l$. Eq. (\ref{eq.hor}) cannot be solved analytically. However, the numerical values of the horizon are given in Table \ref{tab:h1}.

\begin{center}
	\begin{table}[h]
		\begin{center}
			\begin{tabular}{l l r l| r l r l r}
\hline
				\multicolumn{1}{c}{ }&\multicolumn{1}{c}{ $g=0.10$  }&\multicolumn{1}{c}{}&\multicolumn{1}{c|}{ \,\,\,\,\,\, }&\multicolumn{1}{c}{ }&\multicolumn{1}{c}{}&\multicolumn{1}{c}{ $g =0.20$ }&\multicolumn{1}{c}{}\,\,\,\,\,\,\\
				\hline
\multicolumn{1}{c}{ \it{$m$}} & \multicolumn{1}{c}{ $r_-$ } & \multicolumn{1}{c}{ $r_+$ }& \multicolumn{1}{c|}{$\delta$}&\multicolumn{1}{c}{$m$}& \multicolumn{1}{c}{$r_-$} &\multicolumn{1}{c}{$r_+$} &\multicolumn{1}{c}{$\delta$}   \\
				\hline
	\,\,\, 2.85\,\,& \,\,0.145\,\, &\,\,  1.234\,\,& \,\,1.08\,\, &\,\, 2.858&0.253\,\,&\,\,0.843\,\,&\,\,0.590\,\,\\
  	\,\, 2.80\,\, & \,\,0.258\,\, &\,\, 1.073\,\,& \,\,0.815\,\,&2.808&\,\, 0.377\,\,&\,\,0.700\,\,&\,\,0.323\,\,
				\\
	\,\,\, 2.77\,\, &  \,\,0.629\,\,  &\,\,0.629\,\,&\,\,0\,\,&2.77\,& \,\, 0.516\,\,&\,\,0.516\,\,&\,\,0\,\,
				\\ 
				\hline
				\hline
				\multicolumn{1}{c}{ }&\multicolumn{1}{c}{ $m=2.0$  }&\multicolumn{1}{c}{}&\multicolumn{1}{c|}{ \,\,\,\,\,\, }&\multicolumn{1}{c}{ }&\multicolumn{1}{c}{}&\multicolumn{1}{c}{ $m=2.5$ }&\multicolumn{1}{c}{}\,\,\,\,\,\,\\
				\hline
				\multicolumn{1}{c}{ \it{$g$}} & \multicolumn{1}{c}{ $r_-$ } & \multicolumn{1}{c}{ $r_+$ }& \multicolumn{1}{c|}{$\delta$}&\multicolumn{1}{c}{$g$}& \multicolumn{1}{c}{$r_-$} &\multicolumn{1}{c}{$r_+$} &\multicolumn{1}{c}{$\delta$}   \\
				\hline

	\,\,\, 0.20\,\,& \,\,0.145\,\, &\,\,  1.234\,\,& \,\,1.08\,\,&0.20&\,\, 0.253\,\,&\,\,0.843\,\,&\,\,0.590\,\,
				\\
	\,\, 0.22\,\, & \,\,0.258\,\, &\,\, 1.073\,\,& \,\,0.815\,\,&0.22&\,\, 0.377\,\,&\,\,0.700\,\,&\,\,0.323\,\,
				\\
	\,\,\, 0.236\,\, &  \,\,0.629\,\,  &\,\,0.629\,\,&\,\,0\,\,&0.236\,& \,\, 0.516\,\,&\,\,0.516\,\,&\,\,0\,\,
				\\

				\hline
				\hline
    
			\end{tabular}
		\end{center}
		\caption{Inner  horizon ($r_-$), outer horizon ($r_+$) and ($\delta=r_+-r_-$) for different values of magnetic charge $g$ with fixed values of graviton mass  ($m= 2.0$, $m=2.5$), mass $(M=1)$ and $AdS$ length $(l=10)$.}
				\label{tab:h1}
	\end{table}
\end{center}
We plot the horizon's radius for different values of $g$ and $m$  in Fig. \ref{fig:h}.
\begin{figure*}[ht]
\begin{tabular}{c c c c}
\includegraphics[width=.5\linewidth]{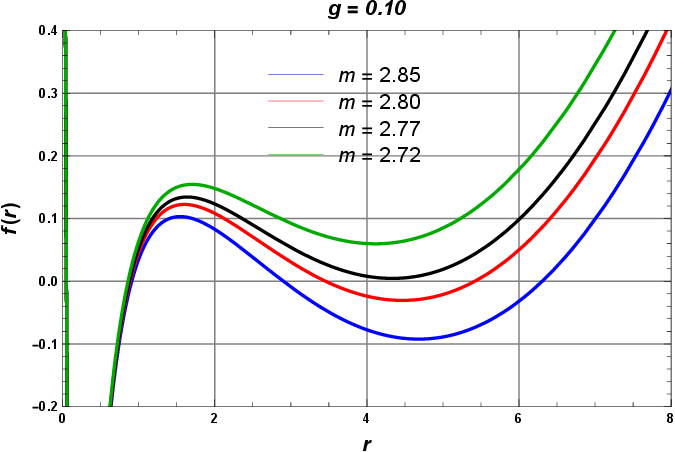}
\includegraphics[width=.5\linewidth]{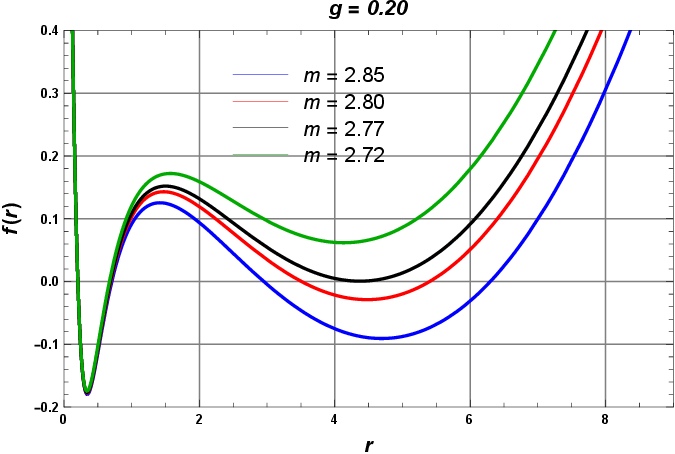}\\
\includegraphics[width=.5\linewidth]{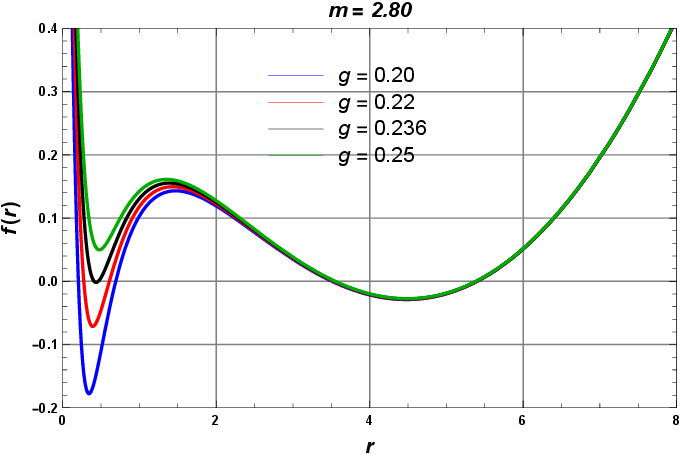}
\includegraphics[width=.5\linewidth]{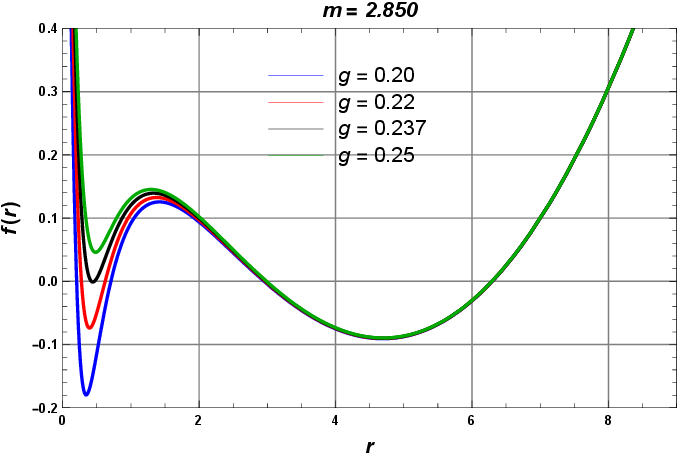}
\end{tabular}
\caption{
The plot illustrates the function \( f(r) \) as a function of the radial coordinate \( r \). The upper panel presents \( f(r) \) for varying values of the graviton mass while keeping the magnetic charge \( g \) fixed. Conversely, the lower panel depicts \( f(r) \) for different values of the magnetic charge \( g \) with a fixed graviton mass.}
\label{fig:h}
\end{figure*}
The parameters $m$ and $g$ have the following effects on the horizon structure of the obtained black hole solution:
\begin{enumerate}
    \item The roots of $f(r)=0$ decide the number of horizons. The obtained black hole solution has four horizons with $g=0.10$ and $m>2.77$ with fixed values of $M=1, c = 0.1,c_1 = -1, c_2 = 1$.
    
    \item The black hole's event horizon radius decreases as the magnetic charge \( g \) and the graviton mass \( m \) increase, indicating a reduction in the black hole's size with increasing values of these parameters.
\end{enumerate}

\begin{figure*}[ht]
\begin{center}
   \includegraphics[width=.65\linewidth]{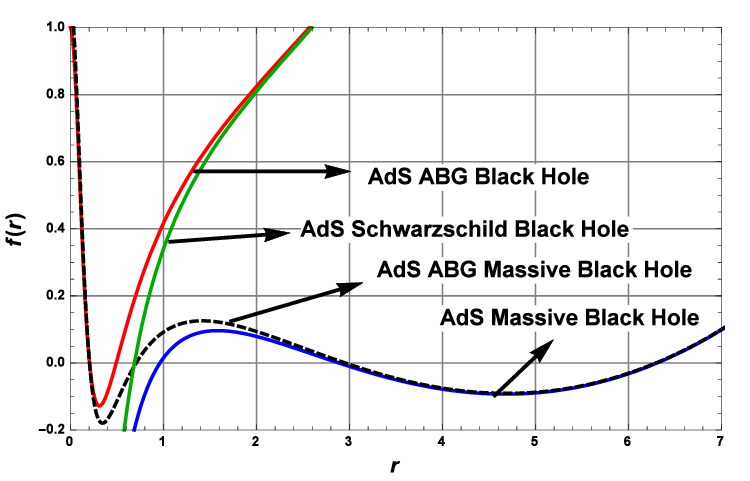} 
\end{center}
\caption{
The plot of the metric function against the radial distance with the limiting values of $m$ and $g$.}
\label{fig:h1}
\end{figure*}

The horizon structure of the $AdS$ ABG massive black hole with limiting cases switching off the magnetic charge ($g=0$), graviton mass ($m=0$) and both ($m=g=0$) are depicted in Fig. \ref{fig:h1}. In the Fig. \ref{fig:h1}, we can see the $AdS$ ABG massive black hole has four horzon and $AdS$ massive black hole has three horzon and they coincide at $r_+> 2.94$ and $AdS$ ABG  black hole has two horizon in contrast with the $AdS$ Schwarzschild black hole, which is coincide at the $r_+> 1.788$. We can also mention that the $AdS$ ABG massive black hole and $AdS$ ABG black hole are coincident until $r_+ < 0.278$.
}

%-------------------------------------------------------------
\section{ Extended Thermodynamics of Black Holes}\label{sec3}
{ The mass of the black hole solution is determined by solving the horizon equation \( f(r)|_{r=r_+} = 0 \), which provides a condition for the event horizon radius and facilitates the extraction of the corresponding mass parameter as
\begin{equation}
M_+=    \frac{(r_+^2+g^2)^{3/2}}{2r_+^2}\left[\left(1+\frac{g^2r_+^2}{(r_+^2+g^2)^2}+\frac{r_+^2}{l^2}\right)+m^2\left(\frac{cc_1 r_+}{2}+c^2c_2\right)\right].
`\end{equation}
{ When the magnetic charge ($m=0$) is absent, the black hole mass simplifies to that of an $AdS$ massive black hole. The limit of vanishing graviton mass corresponds to the $AdS$ ABG black hole mass and the $AdS$ Schawarzschild black hole in the absence of graviton mass and magnetic charge.} Furthermore, asymptotically, the mass transitions to that of an $AdS $massive Reissner-Nordstr\"om black hole. The black hole temperature is computed as follows:
\begin{equation}
    T_+=\left.\frac{f'(r)}{4\pi}\right|_{r=r_+}.  
\end{equation}
We compute the thermodynamic temperature associated with the obtained black hole solution   as
\begin{eqnarray}
T_+&=&\frac{1}{4\pi}\Big[\frac{2r_+}{l^2}+\frac{m^2cc_1}{2}+\frac{2g^2r_+}{g^2+r_+^2}\left(1-\frac{2r_+^2}{g^2+r_+^2}\right)-\nonumber\\&&\qquad\qquad\frac{2g^2-r_+^2}{g^2+r_+^2}\left(1+\frac{g^2r_+^2}{(r_+^2+g^2)^2}+\frac{r_+^2}{l^2}+m^2\left(\frac{cc_1 r_+}{2}+c^2c_2\right)\right)\Big].
\end{eqnarray}
{The temperature of the black hole solution exhibits distinct limiting behaviors: it reduces to the $AdS$ massive black hole temperature when the magnetic charge ($g=0$) is turned off \cite{Cai:2014znn}, to the $AdS$ ABG black hole temperature in the absence of graviton mass ($m=0$),  the $AdS$ Schawarzschild black hole in the absence of ($m=g=0$)} and asymptotically interpolates with the temperature of the AdS massive Reissner-Nordstr\"om black hole.
The dependence of temperature on various parameters are shown in Fig. 
\ref{fig:t}.
\begin{figure*}[ht]
\begin{tabular}{c c c c}
\includegraphics[width=.5\linewidth]{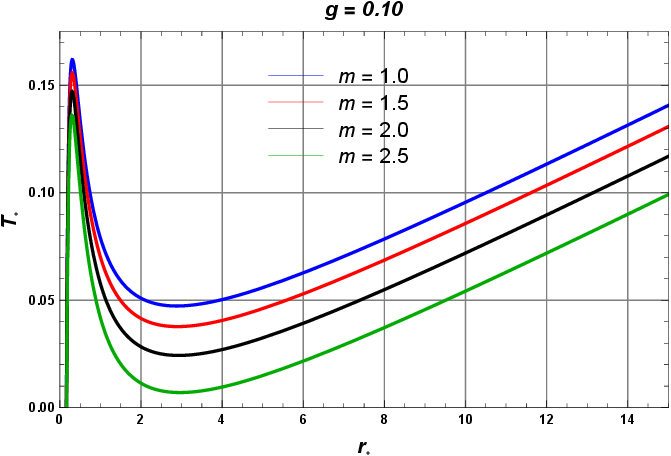}
\includegraphics[width=.5\linewidth]{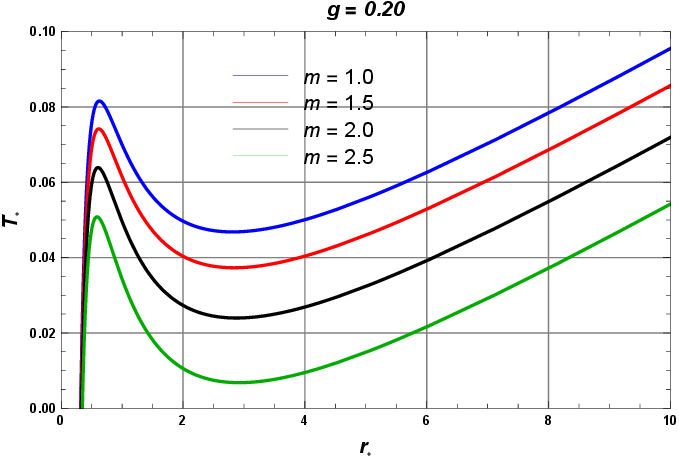}\\
\includegraphics[width=.5\linewidth]{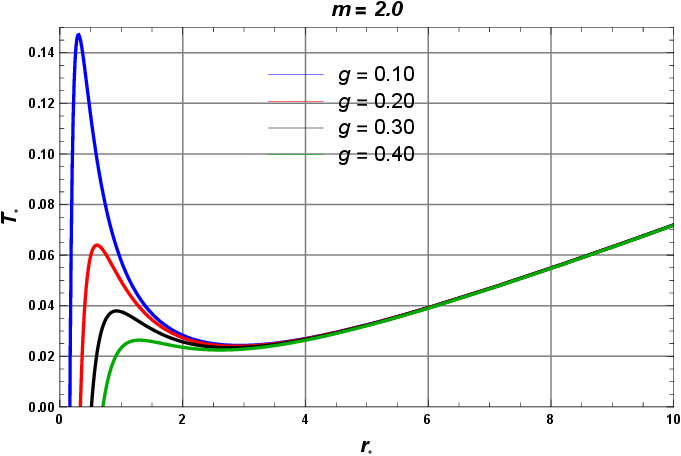}
\includegraphics[width=.5\linewidth]{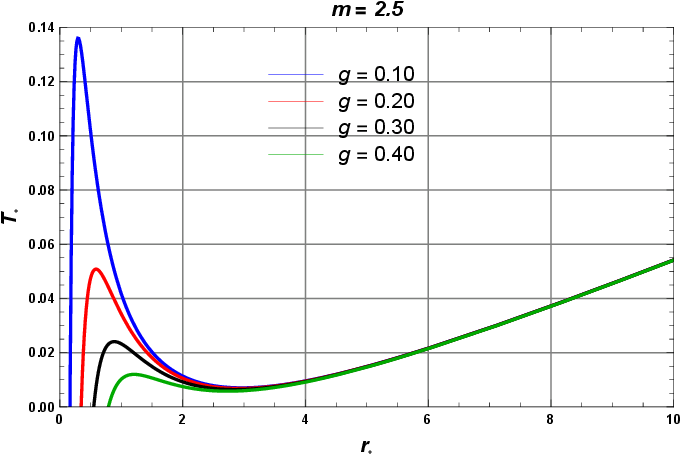}
\end{tabular}
\caption{ The upper panel illustrates the relationship between temperature and horizon radius for various values of graviton mass, while the lower panel shows the temperature vs. horizon radius for different values of magnetic charge, with the graviton mass held constant.}
\label{fig:t}
\end{figure*}
 \begin{center}
\begin{table}[h]
\begin{center}
\begin{tabular}{l|l r l r l| r l r r r}
\hline
\multicolumn{1}{c}{ }&\multicolumn{1}{c}{ }&\multicolumn{1}{c}{ }&\multicolumn{1}{c }{$g=0.1$  }&\multicolumn{1}{c}{  }&\multicolumn{1}{c|}{ }&\multicolumn{1}{c}{  }&\multicolumn{1}{c}{ }&\multicolumn{1}{c }{$g=0.2$}&\multicolumn{1}{c }{} &\multicolumn{1}{c }{}\\
\hline
\,\,$m$\,\, &&  \,\, 1.0\,\, &\,\,1.5\,\, &  \,\,2.0\,\, &\,\,2.5\,\,&&\,\,1.0\,\,&\,\,1.5\,\,&\,\,2.0\,\,&\,\,2.5\,\,
\\  \hline
\,\,$r_{1c} $\,\, &&  \,\, 0.311\,\, &\,\,0.306\,\, &  \,\,0.302\,\, &\,\,0.297\,\,&&\,\,0.996\,\,&\,\,1.088\,\,&\,\,1.36\,\,&\,\,1.39\,\,
\\ 
\,\,$T_+^{1}$\,\,&&\,\,0.162\,\, & \,\,0.157\,\,& \,\,  0.147\,\, &\,\,  0.136\,\,&& \,\,0.095\,\,&\,\, 0.0812\,\,&\,\,0.0786\,\,&\,\,0.0743\,\,
\\ 
\,\,$r_{2c} $\,\, &&  \,\, 2.952\,\, &\,\,2.849\,\, &  \,\,2.712\,\, &\,\,2.746\,\,&&\,\,0.996\,\,&\,\,1.088\,\,&\,\,1.36\,\,&\,\,1.39\,\,
\\ 
\,\,$T_+^{2}$\,\,&&\,\,  0.0467\,\, & \,\,0.0369\,\,& \,\,  0.0246\,\, &\,\,  0.0065\,\,&& \,\,0.095\,\,&\,\, 0.0812\,\,&\,\,0.0786\,\,&\,\,0.0743\,\,
\\
\hline
\end{tabular}
\end{center}
\caption{The variation of the maximum Hawking temperature (\(T_+^{Max}\)) at the critical radius (\(r_c\)) as a function of magnetic charge (\(g\)) and graviton mass (\(m\)), for fixed values of the black hole mass (\(M\)) and length scale parameter (\(l\)).}
\label{tab:temp}
\end{table}
\end{center}
 \begin{center}
\begin{table}[h]
\begin{center}
\begin{tabular}{l|l r l r l| r l r r r}
\hline
\multicolumn{1}{c}{ }&\multicolumn{1}{c}{ }&\multicolumn{1}{c}{ }&\multicolumn{1}{c }{$m=2.0$  }&\multicolumn{1}{c}{  }&\multicolumn{1}{c|}{ }&\multicolumn{1}{c}{  }&\multicolumn{1}{c}{ }&\multicolumn{1}{c }{$m=2.5$}&\multicolumn{1}{c }{} &\multicolumn{1}{c }{}\\
\hline
\,\,$g$\,\, &&  \,\, 0.10\,\, &\,\,0.20\,\, &  \,\,0.30\,\, &\,\,0.40\,\,&&\,\,0.10\,\,&\,\,0.20\,\,&\,\,0.30\,\,&\,\,0.40\,\,
\\  \hline
\,\,$r_{1c} $\,\, &&  \,\, 0.311\,\, &\,\,0.306\,\, &  \,\,0.302\,\, &\,\,0.297\,\,&&\,\,0.996\,\,&\,\,1.088\,\,&\,\,1.36\,\,&\,\,1.39\,\,
\\ 
\,\,$T_+^{1}$\,\,&&\,\,0.162\,\, & \,\,0.157\,\,& \,\,  0.147\,\, &\,\,  0.136\,\,&& \,\,0.095\,\,&\,\, 0.0812\,\,&\,\,0.0786\,\,&\,\,0.0743\,\,
\\ 
\,\,$r_{2c} $\,\, &&  \,\, 2.952\,\, &\,\,2.849\,\, &  \,\,2.712\,\, &\,\,2.746\,\,&&\,\,0.996\,\,&\,\,1.088\,\,&\,\,1.36\,\,&\,\,1.39\,\,
\\ 
\,\,$T_+^{2}$\,\,&&\,\,  0.0467\,\, & \,\,0.0369\,\,& \,\,  0.0246\,\, &\,\,  0.0065\,\,&& \,\,0.095\,\,&\,\, 0.0812\,\,&\,\,0.0786\,\,&\,\,0.0743\,\,
\\
\hline
\end{tabular}
\end{center}
\caption{Dependence of the maximum Hawking temperature (\(T_+^{Max}\)) at the critical radius (\(r_c\)) on the magnetic charge (\(g\)) and graviton mass (\(m\)), with the black hole mass (\(M\)) and length scale parameter (\(l\)) held constant.}
\label{tab:temp}
\end{table}
\end{center}
\begin{figure*}[ht]
\begin{center}
   \includegraphics[width=.65\linewidth]{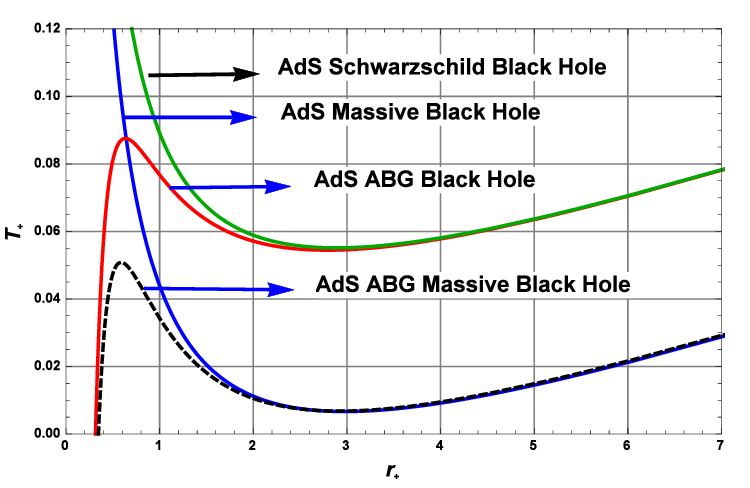} 
\end{center}
\caption{
The Plot of the temperature against the horizon radius with the limiting values of $m$, $g$ and $m,g$.}
\label{fig:h2}
\end{figure*}

The temperature of the $AdS$ ABG massive black hole with limiting cases switching off the magnetic charge ($g=0$), graviton mass ($m=0$) and both ($m=g=0$) are depicted in Fig. \ref{fig:h2}. In the Fig. \ref{fig:h2}, we can see the temperature of $AdS$ ABG massive black hole and $AdS$ massive black hole are coincide at $r_+> 1.687$ and $AdS$ ABG  black hole and  $AdS$ Schwarzschild black hole, which is coincide at the $r_+> 2.413$. We can also mention that the $AdS$ ABG massive black hole and $AdS$ ABG black hole coincide until $r_+ < 0.3991$.

{ Wald derived the first law of thermodynamics in the presence of higher derivative terms
in the action \cite{Wald:1993nt}. The Wald entropy is 
\begin{equation}
S_{W}=\int \frac{\delta S}{\delta R_{\mu\nu \alpha\beta}}\epsilon^{\mu\alpha}\epsilon^{\nu\beta} \sqrt{h} d^2\Omega,\quad \text{with}\quad S=\int d^x \sqrt{-g} {\cal L},
\end{equation}
where $S_W$ is the entropy, $S$ is the action,  $\epsilon^{\mu\nu}$ is the binormal to the horizon, $ h$ is the induced metric on the horizon and $R_{\mu\nu \alpha\beta}$ is the Riemannian tensor is. To calculate the entropy of the obtained black hole solution, we  construct an antisymmetric second-rank tensor $\epsilon_{\mu\nu}$ along these directions so that  $\epsilon_{r t}= \epsilon_{t r}=1$
\begin{equation}
{\cal L}=\frac{1}{16\pi} R_{\mu\nu\alpha\beta}g^{\nu\alpha}g^{\mu\beta},\quad\text{and}\quad \frac{\partial {\cal L}}{\partial  R_{\mu\nu\alpha\beta} }= \frac{1}{16\pi} \frac{1}{2}(g^{\mu\alpha}g^{\nu\beta}-g^{\nu\alpha}g^{\mu\beta}).
\end{equation}
The Wald entropy becomes
\begin{eqnarray}
&&S_W=  \frac{1}{8}\int   \frac{1}{2} (g^{\mu\alpha}g^{\nu\beta}-g^{\nu\alpha}g^{\mu\beta} \epsilon_{\mu\nu}\epsilon^{\alpha\beta} \sqrt{h} d\theta d\phi,\\
&&S_W=  \frac{1}{8}\int 2 \,\, g^{tt}g^{rr}   \sqrt{h}\,\, d\theta d\phi= \frac{1}{4}\int r^2\,\, d\theta d\phi,\\
&&S_W=\pi r_+^2=\frac{A}{4}.
\end{eqnarray}
This entropy agrees with the area law and matches precisely with the entropy of black holes. 
}

Next, we focus on the stability analysis of the black hole, considering both local and global aspects. The sign of the heat capacity determines local stability: a positive heat capacity signifies a stable configuration, whereas a negative value indicates instability. However, we can examine the global stability through the behaviour of the Gibbs free energy. To proceed with this analysis, we now compute the heat capacity of the obtained black hole solution (\ref{bhs}) using the prescribed formula
\begin{equation}
C_+=\frac{\partial M_+ }{\partial T_+}=\bigg(\frac{\partial M }{\partial r_+}\bigg)\bigg(\frac{\partial r_+}{\partial T_+}\bigg).
\label{eq:h}
\end{equation}
 This further simplifies to
\begin{equation}
C_+= \frac{\pi (r_+^2 +g^2)^{5/2}\left(l^2 (A+ m^2 B+) 6g^2r_+^4+12g^2r_+^6+6r_+^8\right)}{9g^6r_+^5+21g^4r_+^7+15g^2r_+^9+3r_+^{11}+l^2r_+(C_++m^2D)},
\label{eq:c}
\end{equation}
where
\begin{eqnarray}
  &&A=-4 g^6 - 6 g^4 r_+^2 -2 g^2 r_+^4 + 2 r_+^6,\\
  &&B=-4  c^2c_2 g^6 -cc_1 g^6 r_+ - 6  c^2c_2 g^4 r_+^2 + 3c c_1 g^2 r_+^5 + 2 c^2c_2 r_+^6 + 2 cc_1 r_+^7,\\
  &&C=2 g^8 + 11 g^6 r_+^2 +12 g^4 r_+^4 +8 g^2 r_+^6 -r_+^8,\\
  &&D= c^2c_2 g^2(2g^6 +11 g^4 r_+^2+5r_+^6+15 g^2r_+^4) +3cc_1 g^6 r_+^3 +6cc_1 g^4 r_+^5  + 3  cc_1 g^2 r_+^7.
\end{eqnarray}
{ This heat capacity exhibits specific limiting behaviours under different conditions. Without the magnetic charge, this is identified with the heat capacity of an $AdS$  massive black hole. Similarly, when the graviton mass is set to zero, it corresponds to the heat capacity of the  $AdS$  ABG black hole and the $AdS$ Schawarzschild black hole in the absence of graviton mass and magnetic charge.} Furthermore, the heat capacity transitions smoothly in the asymptotic regime, interpolating with the massive $AdS$ Reissner-Nordstr\"om black hole. 

\begin{figure*}[ht]
\begin{tabular}{c c c c}
\includegraphics[width=.5\linewidth]{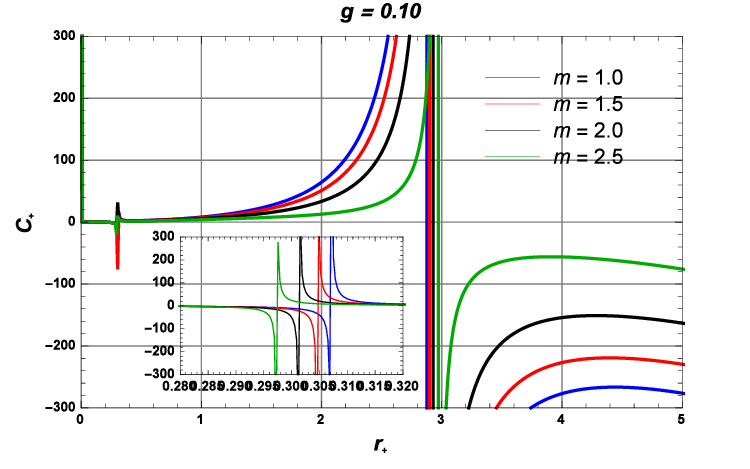}
\includegraphics[width=.5\linewidth]{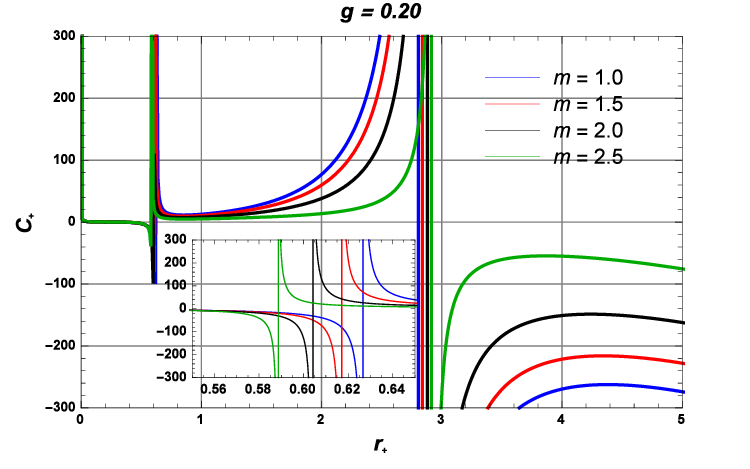}\\
\includegraphics[width=.5\linewidth]{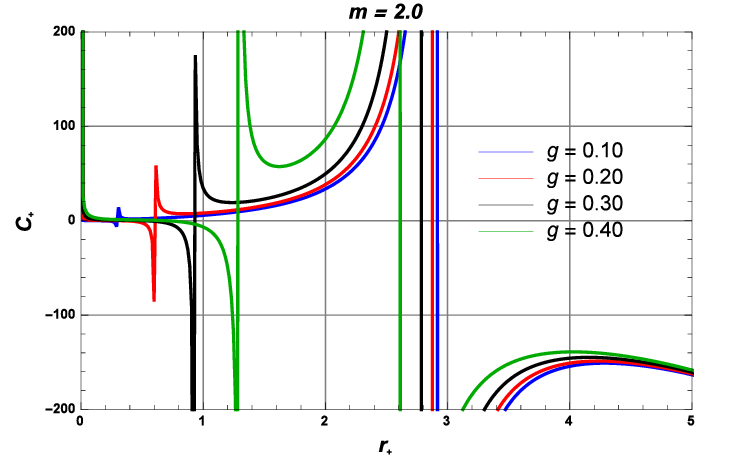}
\includegraphics[width=.5\linewidth]{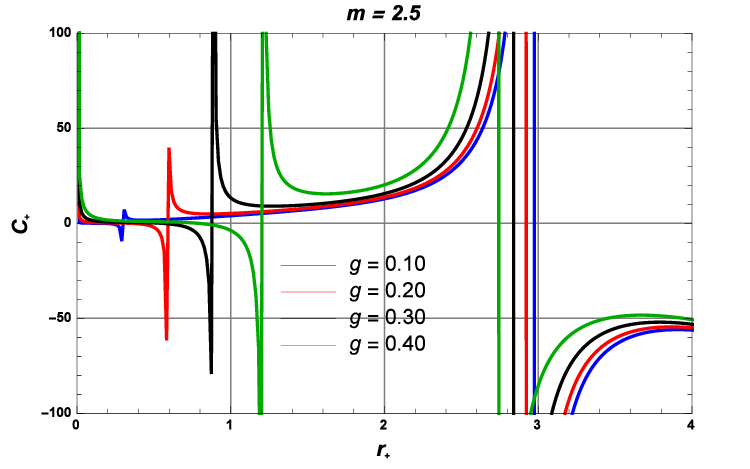}
\end{tabular}
\caption{Plot of heat capacity versus horizon radius: The upper panel shows variations for different values of graviton mass with a fixed magnetic charge, while the lower panel depicts variations for different values of magnetic charge with a fixed graviton mass.}
\label{fig:c}
\end{figure*}

 The stability of the black hole is assessed by analysing the behaviour of the heat capacity \( C \) as a function of the horizon radius \( r_+ \), with variations in the massive gravity parameter \( m \) and Gauss-Bonnet coupling \( \alpha \) (see Fig. \ref{fig:c}). The heat capacity exhibits divergences at critical points \( r_{1+} \) and \( r_{2+} \) (\( r_{1+} < r_{2+} \)), indicating phase transitions. The black hole remains stable for \( r_+ < r_{1+} \) and \( r_+ > r_{2+} \), while it becomes unstable in the intermediate region \( r_{1+} < r_+ < r_{2+} \). These transitions correspond to stability shifts: from stable to unstable at \( r_+ = r_{1+} \) and back to stable at \( r_+ = r_{2+} \). For specific parameter values, the phase transitions occur at \( r_{1+} = 0.48 \) for \( \lambda = 0.1 \) and \( r_{2+} = 1.3 \) for \( \lambda = 0.2 \). The divergence of \( C \) at \( r_+ = r_c \) signifies a second-order phase transition \cite{hp,davis77}. Furthermore, the temperature reaches extrema at the critical points, with maximum \( T_{1+}^{\text{max}} \) at \( r_+ = r_{1+} \) and minimum \( T_{2+}^{\text{max}} \) at \( r_+ = r_{2+} \), reinforcing the correlation between thermal behavior and stability transitions.

\begin{figure*}[ht]
\begin{center}
   \includegraphics[width=.65\linewidth]{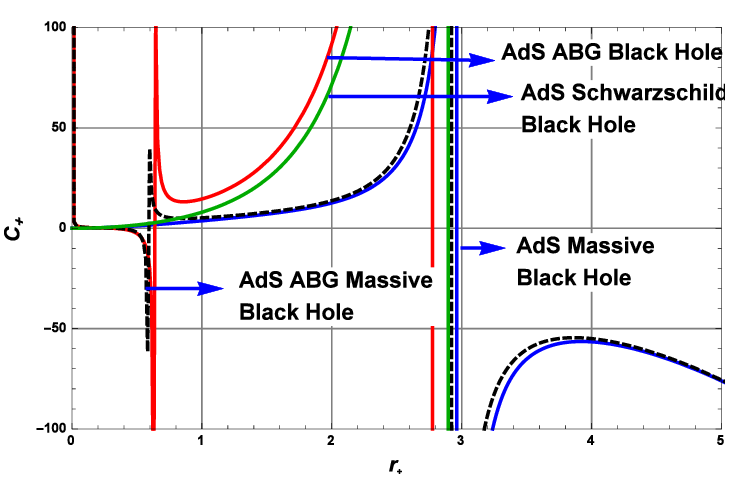} 
\end{center}
\caption{
The Plot of the heat capacity against the horizon radius with the limiting values of $m$, $g$ and $m,g$.}
\label{fig:h9}
\end{figure*}

The heat capacity of the $AdS$ ABG massive black hole with limiting cases switching off the magnetic charge ($g=0$), graviton mass ($m=0$) and both ($m=g=0$) are depicted in Fig. \ref{fig:h9}. In the Fig. \ref{fig:h9}, we can see the heat capacity oof $AdS$ ABG massive black hole and $AdS$ massive black hole are diveges twice at $r_{1+}= 0.5774, \& 0.623$ and $r_{2+}= 2.76, \& 2.94$. The $AdS$ ABG  black hole and  $AdS$ Schwarzschild black hole are diverges at  $r_{+}=2.89, 2.97$.

{Next, we examine the global stability of the black hole, considering the cosmological constant ($\Lambda$) as thermodynamical pressure ($P$), which leads to the interpretation of mass not only as internal energy but also as the Enthalpy of the thermodynamical system. This interpretation leads to the calculation of the Gibbs free energy of the system. The Gibbs free energy serves as a crucial thermodynamic quantity for determining the preferred phase of the black hole. The Gibbs free energy is defined by }
\begin{equation}  
G_+ = H_+-T_+S_+= M - T_+ S_+,  
\end{equation}  
where \( M \) denotes the mass of the black hole, \( T_+ \) represents its Hawking temperature, and \( S_+ \) is the corresponding entropy. The evaluation of this expression enables us to analyse the thermodynamic behaviour and phase structure of the black hole. This gives
\begin{eqnarray}
    G_+&=&  \frac{(r_+^2+g^2)^{3/2}}{2r_+^2}\left[\left(1+\frac{g^2r_+^2}{(r_+^2+g^2)^2}+\frac{r_+^2}{l^2}\right)+m^2\left(\frac{cc_1 r_+}{2}+c^2c_2\right)\right]\nonumber\\&& -r_+^2 \Big[\frac{2r_+}{l^2}+\frac{m^2cc_1}{2}+\frac{2g^2r_+}{g^2+r_+^2}\left(1-\frac{2r_+^2}{g^2+r_+^2}\right)\nonumber\\&&-\frac{2g^2-r_+^2}{g^2+r_+^2}\left(1+\frac{g^2r_+^2}{(r_+^2+g^2)^2}+\frac{r_+^2}{l^2}+m^2\left(\frac{cc_1 r_+}{2}+c^2c_2\right)\right)\Big].
\label{eq:g}
\end{eqnarray}

{The Gibbs free energy exhibits specific limiting behaviours under different conditions. Without the magnetic charge, it reduces to the free energy of an $AdS$ massive black hole. Similarly, when the graviton mass is set to zero, it corresponds to the free energy of the $AdS$ ABG black hole and the $AdS$ Schawarzschild black hole in the absence of graviton mass and magnetic charge.} Furthermore, in the asymptotic limit, the free energy smoothly interpolates with that of the $AdS$ massive Reissner-Nordström black hole.  
We determine the global stability of the black hole by the condition \( G_+ \leq 0 \), which ensures a thermodynamically favoured state. To gain further insights into global stability, we analyse the behaviour of the Gibbs free energy as depicted in Fig. \ref{fig:g}.  

\begin{figure*}[ht]
\begin{tabular}{c c c c}
\includegraphics[width=.5\linewidth]{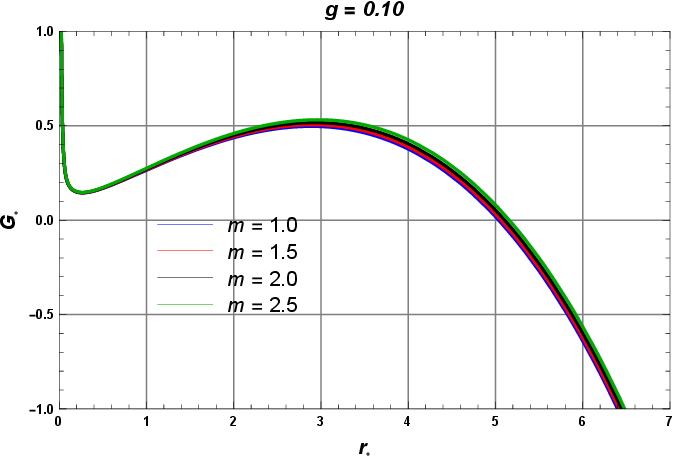}
\includegraphics[width=.5\linewidth]{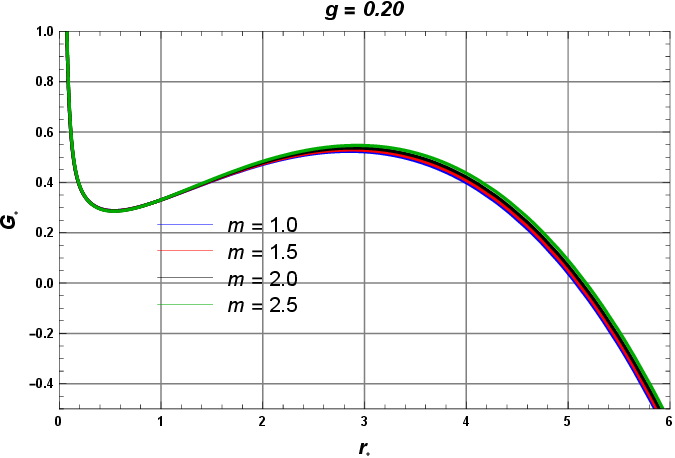}\\
\includegraphics[width=.5\linewidth]{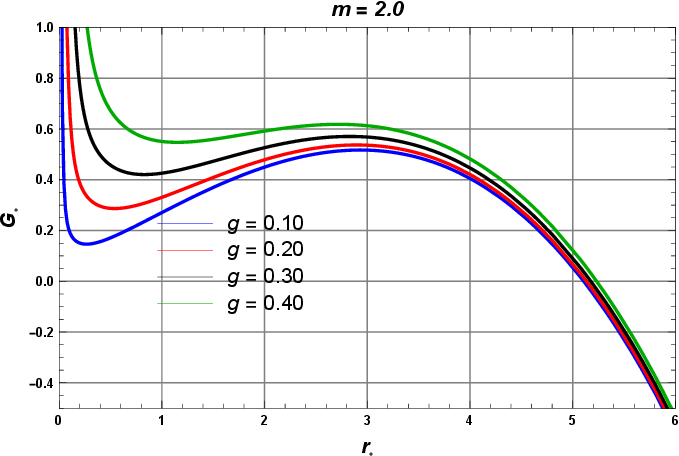}
\includegraphics[width=.5\linewidth]{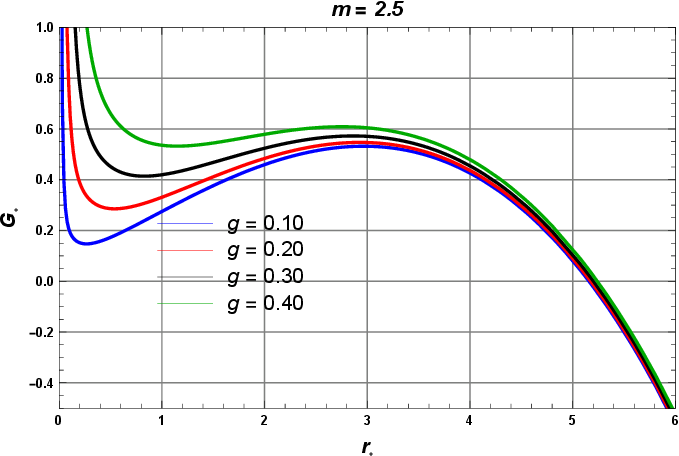}
\end{tabular}
\caption{Plot of Gibbs free energy versus horizon radius: The upper panel illustrates the variation for different values of graviton mass with a fixed magnetic charge, while the lower panel shows the variation for different values of magnetic charge with a fixed graviton mass.}
\label{fig:g}
\end{figure*}
 From the analysis, it is observed that the Gibbs free energy exhibits a local minimum at \( r_{\text{min}} \) and a local maximum at \( r_{\text{max}} \), which are consistent with the extremal points of the Hawking temperature. At these critical points, the nature of the free energy undergoes a transition.  
For \( r_+ < r_{\text{min}} \), the free energy initially decreases, reaching its minimum at \( r_{\text{min}} \). Beyond this point, \( G_+ \) increases with the radius of the horizon until it reaches a maximum at \( r_{\text{max}} \). For \( r_+ > r_{\text{max}} \), the free energy starts to decrease as the horizon radius continues to grow. This behaviour provides valuable insights into the black hole's thermodynamic phase structure and stability.  
\begin{figure*}[ht]
\begin{center}
   \includegraphics[width=.65\linewidth]{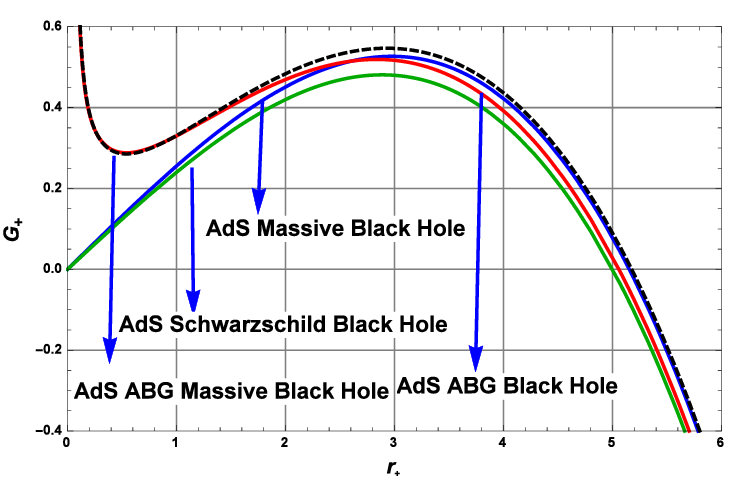} 
\end{center}
\caption{
The Plot of the Gibbs free energy against the horizon radius with the limiting values of $m$, $g$ and $m,g$.}
\label{fig:h10}
\end{figure*}

The Gibbs free energy of the $AdS$ ABG massive black hole with limiting cases switching off the magnetic charge ($g=0$), graviton mass ($m=0$) and both ($m=g=0$) are depicted in Fig. \ref{fig:h10}.
}
%%%%%%%%%%%%%%%%%%%%%%%%%%%%%%%%%%%%%%%%%%%%%%%%%%%%%%%%%%%%
\section{ Photon Radius and Black Hole Shadow}\label{sec4}
Let us analyse the trajectory of a massless particle (photon) propagating in the gravitational field described by the black hole solution (\ref{bhs}). We restrict the photon's motion to the equatorial plane to simplify the analysis by imposing the condition $\theta = \pi/2$. The equations of motion governing this system can be derived from the corresponding Hamiltonian. The Hamiltonian for this system is given by   
\cite{Belhaj:2020okh,Belhaj:2022kek,Belhaj:2021tfc}
\begin{equation}
H=\frac{1}{2}\left[-\frac{p_t^2}{f(r)}+f(r)p_r^2+\frac{p_{\phi}^2}{r^2}\right].
\end{equation}
The canonical conjugate momenta associated with the black hole metric (\ref{bhs}) are derived from the Hamiltonian formulation and can be expressed as follows:
\begin{eqnarray}
&&p_t=\left(1-\frac{2M r^2}{(r^2+g^2)^{3/2}}+\frac{g^2r^2}{(r^2+g^2)^2}+m^2\left (c^2c_2+\frac{cc_1 r}{2}\right)+\frac{r^2}{l^2}\right){\dot t}=E,\\
&&p_r= \left(1-\frac{2M r^2}{(r^2+g^2)^{3/2}}+\frac{g^2r^2}{(r^2+g^2)^2}+m^2\left (c^2c_2+\frac{cc_1 r}{2}\right)+\frac{r^2}{l^2}\right) ^{-1}{\dot r},\\
&&p_{\theta}=r^2{\dot\theta},\\&&p_{\phi}=r^2\sin^2\theta {\dot \phi}=L,
\end{eqnarray}
where $E$ and $L$ are the energy and angular momentum, respectively. These expressions result in the following equations of motion:
\begin{eqnarray}
&&{\dot t}=\frac{E}{ 1-\frac{2M r^2}{(r^2+g^2)^{3/2}}+\frac{g^2r^2}{(r^2+g^2)^2}+m^2\left (c^2c_2+\frac{cc_1 r}{2}\right)+\frac{r^2}{l^2} },\qquad r^2{\dot r}=\pm\sqrt{\cal R},\nonumber\\
&&r^2{\dot \theta}=\pm\sqrt{{ \Theta}}, \qquad  \qquad{\dot\phi}=\frac{L}{r^2},
\end{eqnarray}
where $\cal R$ and $\Theta$, respectively, are  
\begin{eqnarray}
&&{\cal R} = E^2r^4-r^2\left(1-\frac{2M r^2}{(r^2+g^2)^{3/2}}+\frac{g^2r^2}{(r^2+g^2)^2}+m^2\left (c^2c_2+\frac{cc_1 r}{2}\right)+\frac{r^2}{l^2}\right)({\cal K} +L^2),\nonumber\\
&&{\Theta}= {\cal K}-L^2 \cot^2\theta,
\end{eqnarray}
here $\cal K$ is the Carter constant. We can express the radial null geodesic equation as follows: 
\begin{equation}
    {\dot r}^2+V_{eff}(r)=0 \qquad \text{with} \qquad V_{eff}=f(r)\left(\frac{L^2}{r^2}-\frac{E^2}{f(r)}\right).
\end{equation}
 The following equations give the conditions governing null circular geodesics: 
\begin{equation}
V_{eff}=0, \qquad \text{and}\qquad \frac{\partial V_{eff}}{\partial r}=0.
\label{pot}
\end{equation}
These lead to
\begin{equation}
 \sqrt{r^2+g^2}\left(\alpha(g^6 +3g^4r^2  +r^6+3 g^2r^4)+8g^2r^4\right)-M(12g^2r^4+12r^6)|_{r=r_p}=0,
\label{rp}
\end{equation}
where $\alpha=4+ m^2(cc_1r_++4c^2c_2).$

The equation (\ref{rp}) governing the photon radius cannot be solved analytically. However, we obtain numerical solutions for different values of the magnetic charge ($g$) and the graviton mass ($m$), which are presented in Tab. \ref{tr1}. The graphical analysis is depicted in Fig. 
\ref{02}. The results presented in the table indicate that the graviton mass parameter $m$ and the monopole charge $g$ have opposite effects on the photon radius. An increase in graviton mass parameter $m$ increases the photon sphere radius, whereas an increase in the magnetic charge $g$ decreases. This behaviour is consistent with that observed in other black hole spacetimes \cite{Singh:2022dth, Singh:2022ycn}.
It is observed that the graviton mass ($m$) and the monopole charge ($g$) exhibit opposite influences on the photon radius. Specifically, an increase in the graviton mass ($m$) leads to an expansion of the photon sphere, whereas an increase in the magnetic charge ($g$) results in its contraction (see Table \ref{tr}).  
\begin{table}[ht]
 \begin{center}
 \begin{tabular}{ |l | l   | l   | l   |  l |  l | l | l | l | l | }
\hline
            \hline
  \multicolumn{1}{|c}{ } &\multicolumn{1}{c}{}  &\multicolumn{1}{c}{}  &\multicolumn{1}{c }{ }&\multicolumn{1}{c }{$r_p$ }&\multicolumn{1}{c }{ }&\multicolumn{1}{c }{ }&\multicolumn{1}{c |}{ }\\
            \hline
  \multicolumn{1}{|c|}{ $m$} &\multicolumn{1}{c|}{$g=0.1$}  &\multicolumn{1}{c|}{$g=0.2$}  &\multicolumn{1}{c|}{$g=0.3$} &\multicolumn{1}{c|}{$g=0.4$}&\multicolumn{1}{c|}{$g=0.5$}&\multicolumn{1}{c|}{$g=0.6$}&\multicolumn{1}{c|}{$g=0.7$}\\
            \hline
\,\,\,\,\,0.1 ~~  &~~2.992~~  & ~~2.946~~ & ~~2.866~~ & ~~2.745~~& ~~2.568~~& ~~2.295~~& ~~\dots~~ \\            
\,\,\,\,\,0.2~~  &~~3.014~~  & ~~2.970~~ & ~~2.892~~ & ~~2.775~~& ~~2.604~~& ~~2.345~~& ~~1.754~~ \\ 
\,\,\,\,\,0.3 ~~  &~~3.050~~  & ~~3.007~~ & ~~2.933~~ & ~~2.822~~& ~~2.661~~& ~~2.422~~& ~~1.971~~\\ 
\,\,\,\,\,0.4 ~~  &~~3.097~~  & ~~3.057~~ & ~~2.988~~ & ~~2.884~~& ~~2.736~~& ~~2.521~~& ~~2.159~~\\ 
\,\,\,\,\,0.5~~  &~~3.153~~  & ~~3.116~~ & ~~3.052~~ & ~~2.954~~& ~~2.823~~& ~~2.633~~& ~~2.340~~\\ 
            \hline 
\hline
        \end{tabular}
        \caption{The magnitude of photon radius with the variation of graviton mass ($m$) and magnetic charge $g$ for a fixed value of  $M=1$.}
\label{tr}
    \end{center}
\end{table}
{ To analyse the properties of the black hole shadow, we investigate the behaviour of its shadow radius for the black hole solution (\ref{bhs}) in $AdS$ space time. We analyse the
black hole shadow on the observer’s sky. We consider an observer at position ( $r_0,\theta_0$).  We choose an orthonormal tetrad as follows
\begin{eqnarray}
&& e^\mu_{(t)} =\frac{1}{f(r_0)} (1,0,0,0),\qquad e^\mu_{(r)}= f(r_0)(0,1,0,0),\\
&& e^\mu_{(\theta)}=\frac{1}{r_0} (0,0,1,0),\qquad e^\mu_{(\phi)}=\frac{1}{r_0\sin\theta_0} (0,0,0,1).
\end{eqnarray}
The spatial components in the local frame are:
\begin{eqnarray}
 &&p^{(r)}= e^\mu_{(r)}p_\mu = \sqrt{f(r_0)}p_r=\pm E \sqrt{f(r_0)}\sqrt{1-\frac{b^2f(r_0)}{r_0^2}},\\
 && p^{(\phi)}=\frac{L}{r_0}.
\end{eqnarray}
The Celestial angle $\alpha$ is defined as
\begin{equation}
    \tan\alpha=\frac{p^{(\phi)}}{p^{(r)}}.
\end{equation}
For a photon with constants of motion, the conserved quantities $E$, $L$, and the observed angle $\alpha$, the following relations are satisfied. Thus we use 
\begin{equation}
    \sin\alpha=\frac{p^{(\phi)}}{p^{(r)}}=\frac{L}{r_0}\sqrt{\frac{f(r_0)}{E^2}}=\frac{b_c}{r_0}\sqrt{{f(r_0)}}.
\end{equation}
This is the observed angular radius of the shadow for an observer at a finite radius in $AdS$. The shadow radius is:
\begin{equation}
   r_s=r_0\sin\alpha=b_c\sqrt{f(r_0)}=\frac{r_p}{\sqrt{f(r_p)}}\sqrt{f(r_0)}.
\end{equation}
If $r\to\infty$, then $f(r_0)\to 1$, then the asymptotically flat limit does recover. Then the shadow radius of the black hole is reduced to
 \cite{72}
\begin{equation}
r_s=\frac{r_p}{\sqrt{f(r_p)}}.
\end {equation}
}
The numerical values of the impact parameter ($b_c$) are presented in Table \ref{tr1} for different values of the black hole parameters. It is observed that the impact parameter ($b_c$) increases with an increase in the graviton mass ($m$), while it decreases as the magnetic charge ($g$) increases.  
\begin{table}[ht]
 \begin{center}
 \begin{tabular}{ |l | l   | l   | l   |  l |  l | l | l | l | l | }
\hline
            \hline
  \multicolumn{1}{|c}{ } &\multicolumn{1}{c}{}  &\multicolumn{1}{c}{}  &\multicolumn{1}{c }{  }&\multicolumn{1}{c }{$b_c$}&\multicolumn{1}{c }{ }&\multicolumn{1}{c }{ }&\multicolumn{1}{c |}{ }\\
            \hline
  \multicolumn{1}{|c|}{ $m$} &\multicolumn{1}{c|}{$g=0.1$}  &\multicolumn{1}{c|}{$g=0.2$}  &\multicolumn{1}{c|}{$g=0.3$} &\multicolumn{1}{c|}{$g=0.4$}&\multicolumn{1}{c|}{$g=0.5$}&\multicolumn{1}{c|}{$g=0.6$}&\multicolumn{1}{c|}{$g=0.7$}\\
            \hline
\,\,\,\,\,0.1 ~~  &~~3.61~~  & ~~3.59~~ & ~~3.55~~ & ~~3.50~~& ~~3.43~~& ~~3.312~~& ~~\dots~~ \\            
\,\,\,\,\,0.2~~  &~~3.65~~  & ~~3.63~~ & ~~3.59~~ & ~~3.54~~& ~~3.46~~& ~~3.31~~& ~~3.03~~ \\ 
\,\,\,\,\,0.3 ~~  &~~3.72~  & ~~3.70~~ & ~~3.86~~ & ~~3.60~~& ~~3.51~~& ~~3.74~~& ~~3.08~~\\ 
\,\,\,\,\,0.4 ~~  &~~3.83~~  & ~~3.81~~ & ~~4.01~~ & ~~3.70~~& ~~3.60~~& ~~3.45~~& ~~3.17~~\\ 
\,\,\,\,\,0.5~~  &~~3.99~~  & ~~3.97~~ & ~~3.92~~ & ~~3.18~~& ~~3.74~~& ~~3.30~~& ~~3.308~~\\ 
            \hline 
\hline
        \end{tabular}
        \caption{The critical impact parameter ($b_c$) magnitude with the graviton mass variation $m$ and magnetic charge $g$ for a fixed value of  $M=1$.}
\label{tr1}
    \end{center}
\end{table}

\begin{table}[ht]
 \begin{center}
 \begin{tabular}{ |l | l   | l   | l   |  l |  l | l | l | l | l | }
\hline
            \hline
  \multicolumn{1}{|c}{ } &\multicolumn{1}{c}{}  &\multicolumn{1}{c}{}  &\multicolumn{1}{c }{  }&\multicolumn{1}{c }{$r_s$}&\multicolumn{1}{c }{ }&\multicolumn{1}{c }{ }&\multicolumn{1}{c |}{ }\\
            \hline
  \multicolumn{1}{|c|}{ $m$} &\multicolumn{1}{c|}{$g=0.1$}  &\multicolumn{1}{c|}{$g=0.2$}  &\multicolumn{1}{c|}{$g=0.3$} &\multicolumn{1}{c|}{$g=0.4$}&\multicolumn{1}{c|}{$g=0.5$}&\multicolumn{1}{c|}{$g=0.6$}&\multicolumn{1}{c|}{$g=0.7$}\\
            \hline
\,\,\,\,\,0.1 ~~  &~~7.876~~  & ~~7.836~~ & ~~7.765~~ & ~~7.656~~& ~~7.491~~& ~~7.231~~& ~~\dots~~ \\            
\,\,\,\,\,0.2~~  &~~7.863~~  & ~~7.821~~ & ~~7.747~~ & ~~7.631~~& ~~7.427~~& ~~7.186~~& ~~6.550~~ \\ 
\,\,\,\,\,0.3 ~~  &~~7.844~  & ~~7.983~~ & ~~8.151~~ & ~~7.596~~& ~~7.410~~& ~~7.114~~& ~~6.513~~\\ 
\,\,\,\,\,0.4 ~~  &~~7.820~~  & ~~7.771~~ & ~~8.190~~ & ~~7.553~~& ~~7.352~~& ~~7.042~~& ~~6.471~~\\ 
\,\,\,\,\,0.5~~  &~~7.781~~  & ~~7.743~~ & ~~7.651~~ & ~~8.156~~& ~~7.296~~& ~~6.445~~& ~~6.454~~\\ 
            \hline 
\hline
        \end{tabular}
        \caption{The critical shadow radius ($r_s$) magnitude with the graviton mass variation $m$ and magnetic charge $g$ for a fixed value of  $M=1$.}
\label{tr1}
    \end{center}
\end{table}

\begin{figure*}[ht]
\begin{tabular}{c c c c}
\includegraphics[width=.45\linewidth]{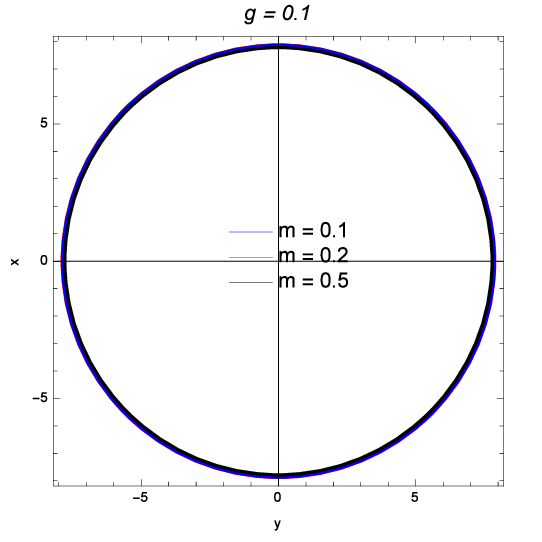}
\includegraphics[width=.45\linewidth]{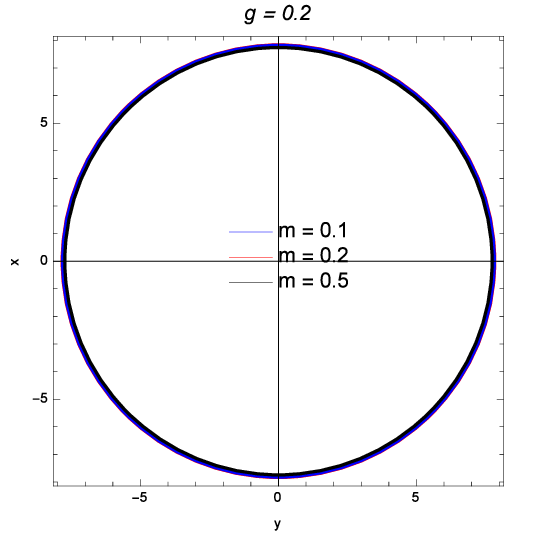}\\
\includegraphics[width=.45\linewidth]{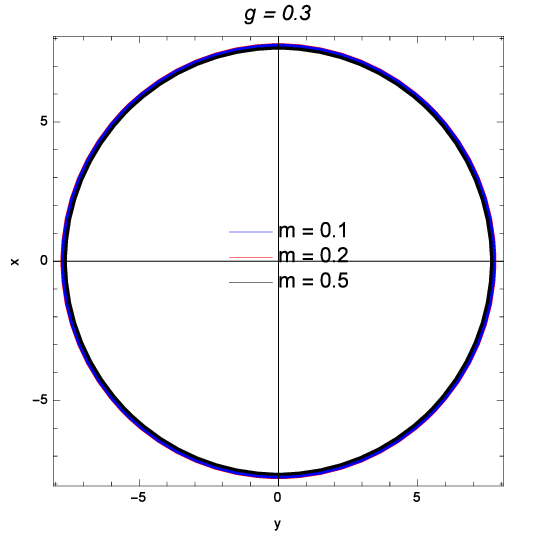}
\includegraphics[width=.45\linewidth]{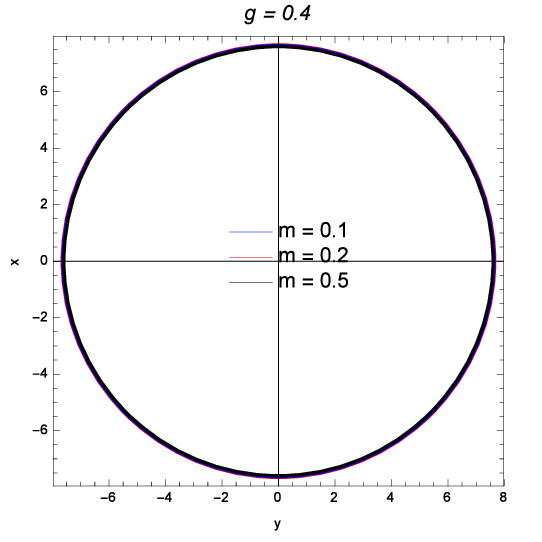}
\end{tabular}
\caption{The shadow of the black hole for different values of the graviton mass ($m$) and magnetic monopole charge ($g$), specifically for $m, g = 0.1,\, 0.2,\, 0.3,$ and $0.5$, with the black hole mass fixed at $M = 1$.
}
\label{fig02}
\end{figure*}

 %----------------------------------------------------------------------------------------------------------
\section{Quasinormal Modes}\label{sec5}
{To investigate the dynamical stability of the black hole solution, we analyse its QNMs. Complex frequencies characterise these modes, expressed as $\omega = \omega_R + i \omega_I$, where $\omega_R$ and $\omega_I$ denote the real and imaginary components, respectively. The sign of $\omega_I$ determines the stability of the black hole: if $\omega_I > 0$, the perturbations grow exponentially, indicating an instability, whereas if $\omega_I < 0$, the perturbations decay, signifying stability. The  QNMs and quasinormal frequencies (QNFs) can be obtained by solving the scalar field equation in the black hole spacetime \cite{Singh:2022xgi}
%------------------------------------------------------------n
\begin{equation}
\frac{1}{\sqrt{-g}}\partial_{\mu}\left(\sqrt{-g} g^{\mu\nu}\partial_{\nu}\right)\phi=0.
\label{scalar1}
\end{equation}
The solution can be derived using the method of separation of variables, yielding
\begin{equation}
\phi=\frac{1}{r}\sum_{lm}e^{i\omega t} u_{lm}(r)Y^m_{l}(\theta,\phi),
\label{scalar2}
\end{equation}
where $Y^m_{l}$ represent the spherical harmonics. By introducing the tortoise coordinate, defined as $dr^{*} = dr / f(r)$, the radial equation can be recast into a Schrödinger-like form 
\begin{equation}
\left(\frac{d^2}{dr^{*^2}}+\omega^2-V_0(r^{*})\right) u(r)=0,
\end{equation}
where 
\begin{equation}
V_0(r^{*})=f\left(\frac{f'}{r}+\frac{l(l+1)}{r^2}\right).   
\end{equation}
The WKB quantisation condition
 We plot the effective potential for different values of the harmonic index $l$ with fixed values of graviton mass and magnetic monopole charge  ($m,g$). When $l$ is increased, the potential height increases. There is a local minimum between the Cauchy and the event horizon when. When the deviation 

To find the QNF, one has to impose boundary conditions near the event horizon. These boundary conditions can be written as

\begin{eqnarray}
   && \phi(r_\star)\to e^{i\omega r}, \qquad\qquad r\to -\infty,\nonumber\\
    && \phi(r_\star)\to e^{-i\omega r}, \qquad\qquad r\to \infty,\nonumber
\end{eqnarray}
where the $\pm$ sign corresponds to ingoing waves at the horizon and outgoing waves at infinity. The frequencies corresponding to the QNM are given by $\omega_R+i\omega_I$, where $\omega_R$ and $\omega_I$ are the oscillating damping components of the frequency. Since the effective potential has a peak and has the form of a potential barrier (see Fig. ), to find the QNM frequencies, one can use the WKB approach, which was first developed by Schutz and Will \cite{will}  and extended to third order \cite{iyer}, and then extended to sixth order \cite{konoplya1}. The WKB formula has the form
\begin{equation}
   \frac{ i(\omega^2-V_0)}{\sqrt{-V_0''}}=-\sum_{n=2}^k \Lambda_k=n+\frac{1}{2},
\end{equation}
where $V_0$ is the maximum height of the potential and $V_0$ is its second derivative with respect to the tortoise coordinate evaluated at the radius where $V_0$ reaches a maximum. The values $\Lambda_k$ are corrections that depend on the value of the potential and higher derivatives of it at the maximum. The exact expressions for the terms $\Lambda_k$ are too long \cite{konoplya1}. Now, we will discuss the properties of QNF for massless scalar fields for $l>0$ modes. The plot of real and imaginary values of QNM is depicted in Tab. \ref{tab:q2}.

\begin{table}[ht]
 \begin{center}
 \begin{tabular}{ l | l    l    l   |  l |  l | l      }
\hline
           
            \hline
  \multicolumn{1}{l|}{$l$} &\multicolumn{1}{l}{$\omega_R$}  &\multicolumn{1}{l}{$\omega_I$}  \\ 
            \hline
 1    & 0.431   & 0.163   \\            
 2  & 0.712  & 0.158  \\ 
3   & 0.901  & 0.144 \\ 
            \hline 
        \end{tabular}
    \caption{ The  real and imaginary values of QNM using the WKB, different values of quantum numbers $l = 1,2,3$ and $n = 0$ with fixed value of $M = 1$.}
\label{tab:q2}
    \end{center}
\end{table}

%%%%%%%%%%%%%%%%%%%%%%%%%%%%%%%%%%%%%%%%%%%%%%%%%%%%%%%%%%%%%%%%%%%%%%%%%%%
\subsection{Quasinormal modes in Eikonal limit}
In the eikonal (large $l$) limit, the WKB approximation \cite{will, iyer, konoplya1, wkb1, yar} is employed to determine the QNFs, expressed as  
\begin{equation}
\omega=l\Omega-i\left(n+\frac{1}{2}\right)|\Lambda|,    
\end{equation}
 with    
\begin{eqnarray}
 \Omega=\frac{\sqrt{f(r_p)}}{r_p}=\frac{1}{L_p}\qquad \text{and}\qquad \Lambda=\frac{\sqrt{2f(r_p)-r^2_pf''(r_p)}}{\sqrt{2} L_p}.
\end{eqnarray}
\begin{figure*}[ht]
\begin{tabular}{c c c c}
\includegraphics[width=.5\linewidth]{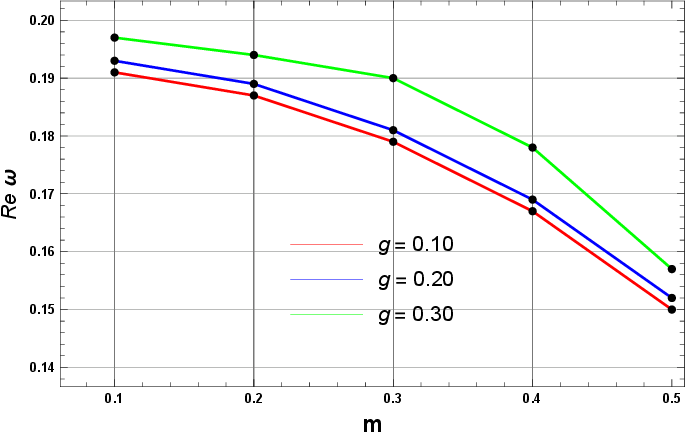}
\includegraphics [width=.5\linewidth]{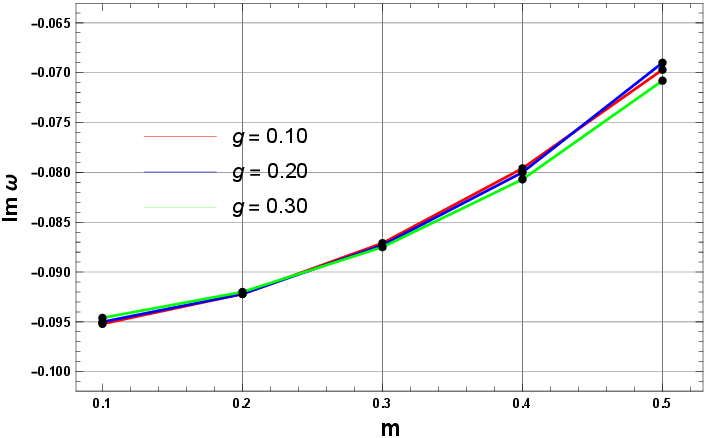}
\end{tabular}
\caption{The real part (left panel) and imaginary part (right panel) of the  QNM frequencies as a function of the magnetic monopole charge ($g$) for a fixed black hole mass ($M$).  
 }
\label{fig:q}
\end{figure*}

The numerical values of the real and imaginary parts of the  QNFs are provided in Table \ref{tab:q} and graphically represented in Fig. \ref{fig:q} for various values of the black hole parameters. The negativity of the imaginary part of the  QNMs indicates that the perturbations decay over time, confirming the stability of the black hole solution.

The influence of the parameters ($m, g$) on the behaviour of  QNMs and  QNFs is illustrated in Fig. \ref{fig:q}. It is observed that the real part of the fundamental QNM decreases with increasing ($m$). In contrast, the imaginary part initially increases (indicating reduced damping) sharply and subsequently increases more slowly, approaching an approximately constant value for larger ($m$).  
\begin{table}[ht]
 \begin{center}
 \begin{tabular}{ l | l   | l   | l   |  l |  l | l | l     }
\hline
            \hline
  \multicolumn{1}{c}{ } &\multicolumn{1}{c}{}  &\multicolumn{1}{c}{}  &\multicolumn{1}{c }{  }&\multicolumn{1}{c }{$\omega_R + i\, \omega_I$}&\multicolumn{1}{c }{ }&\multicolumn{1}{c }{ }&\multicolumn{1}{c }{ }\\
            \hline
  \multicolumn{1}{l|}{$m$} &\multicolumn{1}{l|}{$g=0.1$}  &\multicolumn{1}{l|}{$g=0.2$}  &\multicolumn{1}{l|}{$g=0.3$} &\multicolumn{1}{l|}{$g=0.4$}&\multicolumn{1}{l|}{$g=0.5$}&\multicolumn{1}{l|}{$g=0.6$}&\multicolumn{1}{l }{$g=0.7$}\\ 
            \hline
 .1    & .277-.137 i  & .278-.136 i & .280-.134 i & .285-.131 i& .291-.125 i& .301-.114 i & ~~\dots~~ \\            
 .2  & .273-.135 i & .275-.134 i & .278-.129 i & .282-.129 i& .288-.124 i& .300-.115 i& .329-.064 i \\ 
 .3   & .268-.130 i & .270-.129 i & .258-.121 i & .277-.126 i& .284-.122 i& .296-.115 i& .323-.090 i\\ 
 .4  & .260-.124 i   & .262-.123 i & .249-.112 i& .270-.121 i& .277-.118 i&  .289-.113 i & .315-.099 i\\ 
 .5   & .250-.115i & .251-.114 i & .254-.114 i & .239-.100 i& .267-.111 i& .302-.081 i & .302-.010 i\\ 
            \hline 
\hline
        \end{tabular}
    \caption{Numerical values of the real part of the  QNMs for different values of the graviton mass ($m$) and magnetic charge ($g$), with the black hole mass fixed at $M = 1$, and quantum numbers $l = 1$ and $n = 0$.  
}
\label{tab:q}
    \end{center}
\end{table}

%\section{Comparison between WKB and Eikonal} 
The Comparison between WKB and eikonal is tabulated in the Tab. \ref{tab:q1} using the WKB and eikonal for different values of quantum numbers $l = 1,2,3$ and $n = 0$ with a fixed value of $M = 1$.

\begin{table}[ht]
 \begin{center}
 \begin{tabular}{ l | l   | l   | l   |  l |  l | l      }
\hline
  \multicolumn{1}{l|}{$l$} &\multicolumn{1}{l|}{$\omega_R^{WKB}$}  &\multicolumn{1}{l|}{$\omega_R^{Eik}$}  &\multicolumn{1}{l|}{$\omega_R^{WKB}/\omega_R^{Eik}$} &\multicolumn{1}{l|}{$\omega_I^{WKB}$}&\multicolumn{1}{l|}{$\omega_I^{Eik}$}&\multicolumn{1}{l}{$\omega_I^{WKB}/\omega_I^{Eik}$}\\ 
            \hline
 1    & 0.431  & 0.277 & 1.55 & 0.163 i& 0.137 & 1.189  \\            
 2  & 0.712 & 0.547 & 1.301 & 0.158 i& 0.135 &  1.170 \\ 
3   & 0.901 & 0.805 & 1.11 & 0.144 & 0.130 & 1.107\\ 
            \hline 
\hline
        \end{tabular}
    \caption{ The comparison between the numerical values of  real and imaginary values of QNM using the WKB and eikonal for different values of quantum numbers $l = 1,2,3$ and $n = 0$ with fixed value of $M = 1$.}
\label{tab:q1}
    \end{center}
\end{table}
In the table \ref{tab:q1}, we can see that the real part $\omega_R$ for $l$ deviates substantially from the eikonal value. The deviation decreases as $l$ increases for real and  imaginary part $\omega_I$ part of the QNMs to the eikonal value even for small $l$. So we can say that the $\omega_R$ converges more slowly to the eikonal $l\Omega$, whereas $\omega_I \neq (n+1/2)\Lambda$.}
%-------------------------------------------------------------------
\section{Results and Conclusions}\label{sec6}
 In this study, we have analysed the dynamic behaviour and structure of a black hole with a graviton mass and a magnetic charge. The Gibbs free energy had exhibited specific limiting behaviors under different conditions, reducing to the free energy of an AdS massive black hole in the absence of magnetic charge and to that of an $AdS $ABG black hole when the graviton mass had been set to zero. In the asymptotic limit, the free energy had smoothly interpolated with that of the $AdS$ massive Reissner-Nordström black hole. The stability of the black hole had been determined by the condition \( G_+ \leq 0 \), ensuring a thermodynamically favoured state.  

Furthermore, we have examined the photon sphere and black hole shadow. It had been observed that an increase in the graviton mass had led to an expansion of the photon sphere and shadow radius, whereas an increase in the magnetic charge had resulted in their contraction. These results are consistent with those previously studied in black hole spacetimes.  

Also, we have investigated the  QNMs to explore the dynamical stability of the black hole. The numerical results have shown that the imaginary part of the QNMs remains negative, confirming that the perturbations decay over time, thereby indicating the stability of the black hole solution. Additionally, the real part of the QNM frequency has decreased, while the imaginary part has initially increased, stabilising to significant values. These findings have provided valuable insights into the interplay between graviton mass, magnetic charge, and black hole stability, contributing to a deeper understanding of modified gravity theories.

\section*{Data Availability Statement} 
Data sharing does not apply to this article, as no data sets were used or analysed during the current study.

\section*{Acknowledgements}
This research was funded by the Science Committee of the Ministry of Education and Science of the Republic of Kazakhstan (Grant No. AP22682760).
  DVS would like to thank the Council of Science and Technology, Uttar Pradesh, for the project (grant no. CST/D-828).

 \appendix
 \section{{Dimensional analysis of massive parameters}}
 The dimensional interpretation of massive terms is
 \begin{eqnarray}
   c=[L],\ \  c_1=c_2= [M^{-2}L^{-2}],\ \  c^2c_2 =[M^{-2}],\ \ cc_1 =[M^{-1}].
 \end{eqnarray}

 \section{Mathematica Codes for Plots}
 The Mathematica codes of thermodynamic quantities associated with the black hole are depicted below for $c_1=-1, c_2=1$  and $c=0.1$ 
\begin{figure*}[hbt]
\begin{tabular}{c c c c}
\includegraphics[width=1.1\linewidth]{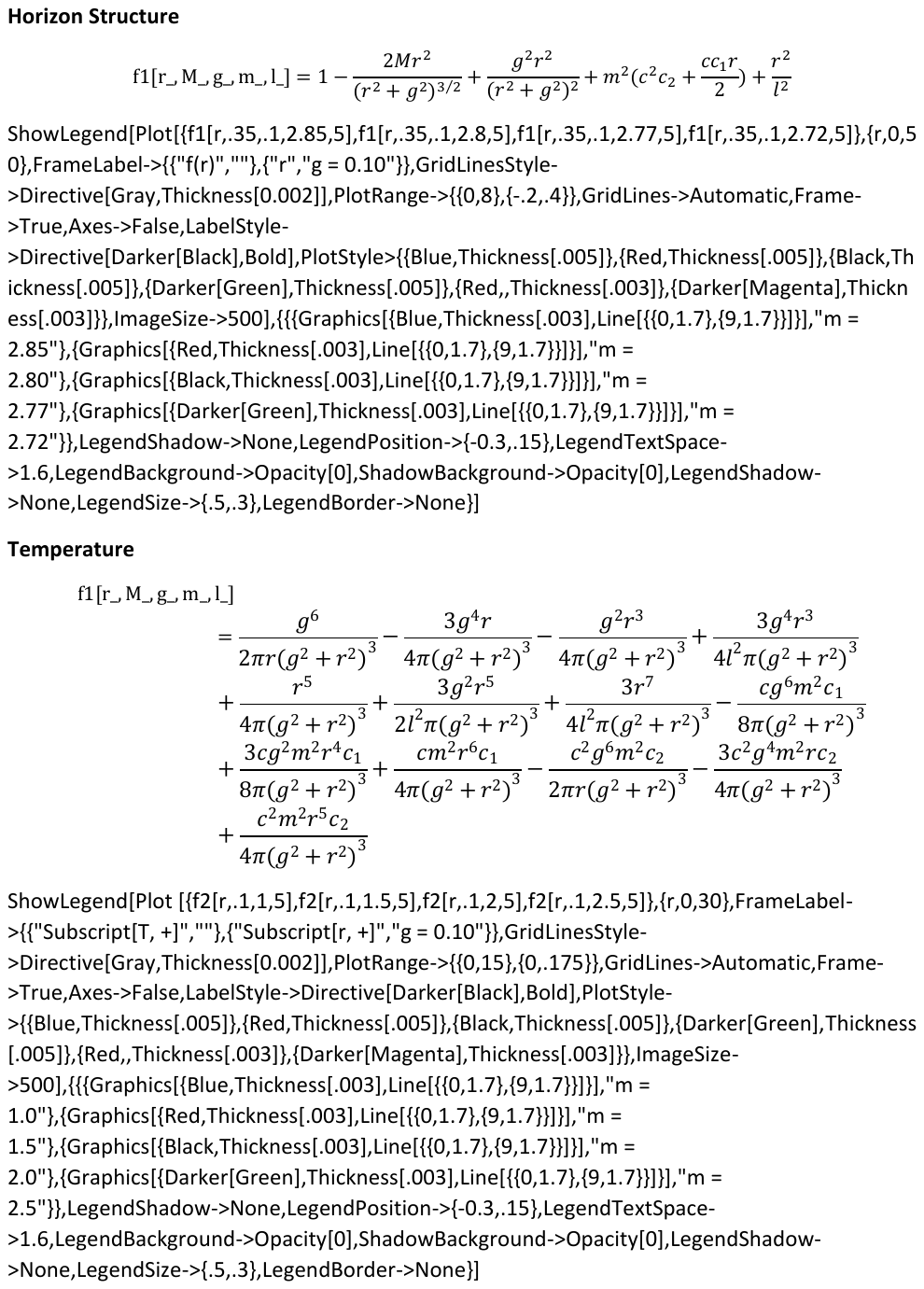}
\end{tabular} 
\end{figure*}
\begin{figure*}[hbt]
\begin{tabular}{c c c c}
\includegraphics[width=1.1\linewidth]{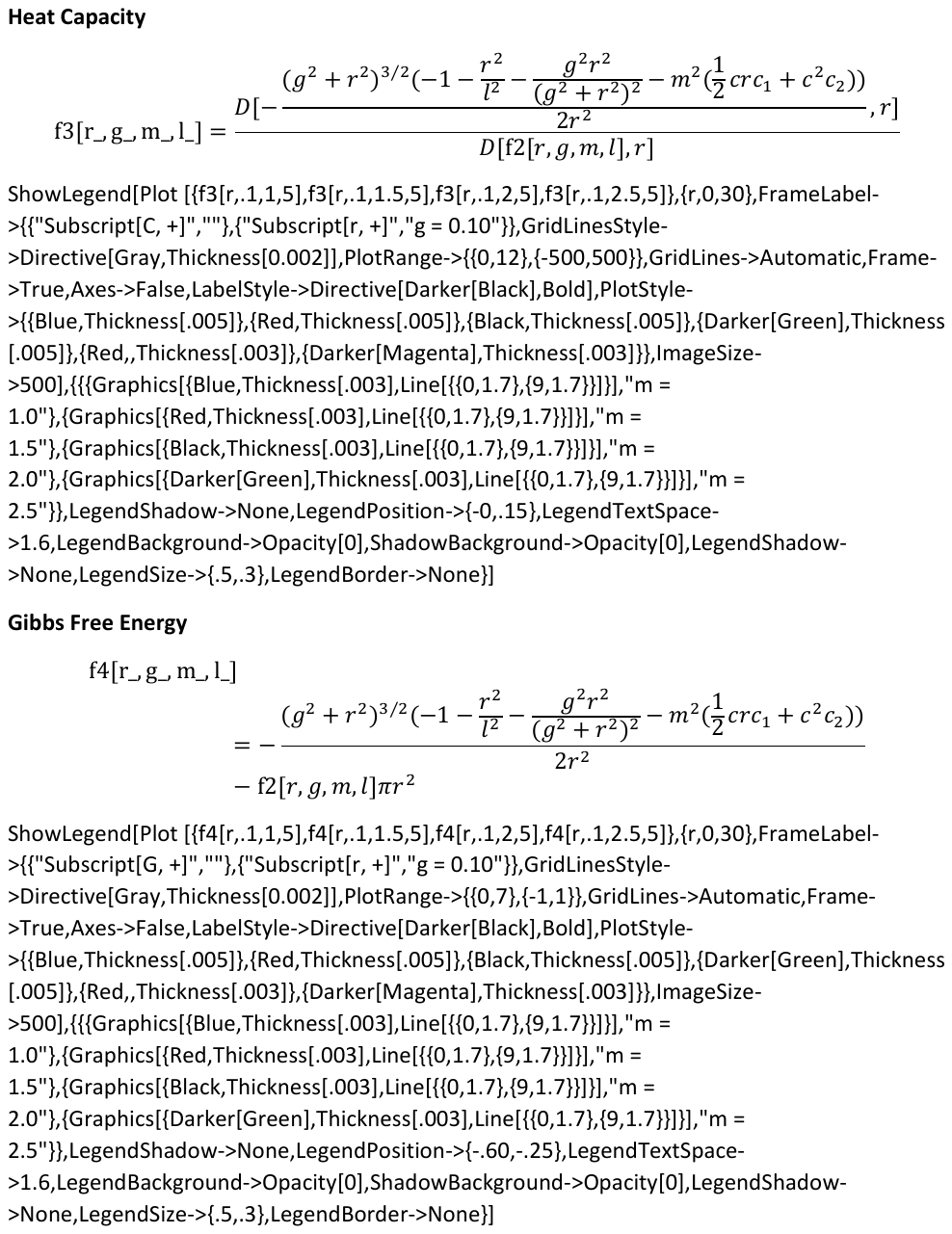}
\end{tabular}  
\end{figure*}
\end{document}